\documentclass[twocolumn, twocolappendix]{aastex631}

\usepackage{amsmath}
\usepackage{mathrsfs}

\def\bB{\boldsymbol{B}}

\begin{document}
\nolinenumbers
\title{Origin of Pulsed Radio Emission from Magnetars}

\author[0009-0002-0127-687X]{Shuzhe Zeng}
\affiliation{
 Department of Physics, University of Maryland, College Park, MD 20742, USA
}
\affiliation{
 Institute for Research in Electronics and Applied Physics, University of Maryland, College Park, MD 20742, USA
}

\author[0000-0001-7801-0362]{Alexander Philippov}
\affiliation{
 Department of Physics, University of Maryland, College Park, MD 20742, USA
}
\affiliation{
 Institute for Research in Electronics and Applied Physics, University of Maryland, College Park, MD 20742, USA
}

\author[0000-0001-6835-273X]{James Juno}
\affiliation{Princeton Plasma Physics Laboratory, Princeton, NJ 08543, USA}

\author[0000-0001-5660-3175]{Andrei M. Beloborodov}
\affiliation{Physics Department and Columbia Astrophysics Laboratory, Columbia University, 538 West 120th Street New York, NY 10027,USA}
\affiliation{Max Planck Institute for Astrophysics, Karl-Schwarzschild-Str. 1, D-85741, Garching, Germany}

\author{Elena Popova}
\affiliation{
 Department of Physics, University of Maryland, College Park, MD 20742, USA
}

\begin{abstract}
Extended periods of radio pulsations have been observed for six magnetars, displaying characteristics different from those of ordinary pulsars. In this Letter, we argue that radio emission is generated in a closed, twisted magnetic flux bundle originating near the magnetic pole and extending beyond 100\,km from the magnetar. The electron-positron flow in the twisted bundle has to carry electric current and, at the same time, experiences a strong drag by the radiation field of the magnetar. This combination forces the plasma into a ``radiatively locked'' state with a sustained two-stream instability, generating radio emission. We demonstrate this mechanism using novel first-principles simulations that follow the plasma behavior by solving the relativistic Vlasov equation with the discontinuous Galerkin method. First, using one-dimensional simulations, we demonstrate how radiative drag induces the two-stream instability, sustaining turbulent electric fields. When extended to two dimensions, the system produces electromagnetic waves, including superluminal modes capable of escaping the magnetosphere. We measure their frequency and emitted power, and incorporate the  local simulation results into a global magnetospheric model. The model explains key features of observed radio emission from magnetars: its appearance after an X-ray outburst, wide pulse profiles, luminosities $\sim 10^{30}{\rm{erg/s}}$, and a broad range of frequencies extending up to $\sim 100\, \mathrm{GHz}$.
\end{abstract}


\section{Introduction} \label{sec:intro}

The remarkable magnetospheric
activity of magnetars likely results from
their surface deformations driven by internal magnetic
stresses \citep{kaspi&beloborodov_2017}. These motions twist the magnetospheric field lines, creating $\nabla \times \boldsymbol{B} \parallel \bB$ in the nearly force-free magnetosphere \citep{Thompson_2002,Beloborodov_2007}. The twist of magnetic field $\bB$ is sustained by electric current $\boldsymbol{j}_0 = (c/4\pi)\nabla \times \boldsymbol{B}$, which has a long lifetime (1-10\,yr) on field lines that extend far from the magnetar \citep{Beloborodov_2009}. The gradual resistive decay of the magnetospheric twist explains the pulsating hard X-ray emission from magnetars \citep{Beloborodov_2013a,Hascoet_2014} and likely also powers their radio pulsations. 

Among the 30 known magnetars, six have been observed to enter extended periods of radio pulsations \citep{Olausen_2014}.\footnote{\url{https://www.physics.mcgill.ca/~pulsar/magnetar/main.html}} 
They exhibit properties substantially different from those of ordinary radio pulsars. First, the width of observed radio pulses is significantly larger than the small angular size of the open field line bundle \cite[e.g.][]{Levin_2010}, sometimes covering a substantial fraction of the full rotational period. Second, 
the radio power of magnetars, $L_r \sim 10^{30}\,{\rm{erg/s}}$, exceeds that of ordinary radio pulsars by about 2 orders of magnitude
\cite[e.g.][]{Camilo_2006}. Third, the radio spectrum 
significantly evolves and can be unusually hard, extending to frequencies as large as $\sim 100{\rm GHz}$ \cite[e.g.][]{Camilo_2008, Torne_2022}. Fourth, the vast majority of radio emission events occur after X-ray outbursts 
and last up to a few years
\cite[e.g.][]{Esposito_2020}. 

Two proposals have been put forward to explain the long-lived radio pulsations from
magnetars:

(1) The emission is
powered by enhanced dissipation near the separatrix between the open and closed twisted field lines,
which extends to the light cylinder $R_{\rm LC}=cP/2\pi\sim 10^{10}$\,cm \citep{Thompson_2008}, where $P=2\pi/\Omega$ is the rotation period of the magnetar. This picture invokes an electric current 
$I_{\rm LC}\sim \mu\Omega^2/c$, where $\mu\sim B_*R^3_*/2$ is the magnetospheric dipole moment, $B_*\sim 10^{15}{\rm G}$ is the magnetic field strength at the magnetar pole, and $R_*\sim 10\,{\rm km}$ is the stellar radius. Powering the observed radio emission by 
the dissipation rate 
\begin{equation}
    I_{\rm LC}\Phi \sim 9\times 10^{28} {\rm{erg/s}}\left(\frac{\Phi} {\mathrm{GV}}\right)\left(\frac{B_*}{10^{15}\mathrm{G}}\right)\left(\frac{P}{5\mathrm{s}}\right)^{-2},
\end{equation}
with a reasonable efficiency requires a high voltage $\Phi$, far exceeding the typical  $\Phi \sim 1$\,GV that sustains electric currents in the inner magnetosphere \citep{Beloborodov_2007}. An increase of $\Phi$ toward the light cylinder would also lead to a quick resistive untwisting of the outer magnetosphere \citep{Beloborodov_2009}, which makes it challenging to explain the long lifetime of radio emission.

(2) The emission comes from magnetic field lines closing well inside the light cylinder \citep{Beloborodov_2013}. This proposal is based on the realization that the plasma carrying magnetospheric currents is forced into a peculiar ``radiatively-locked'' state, which must sustain a two-stream instability. \cite{Beloborodov_2013} conjectured that this instability generates radio emission, however did not demonstrate the generation mechanism and its efficiency.

In this Letter, we quantitatively explore the proposal of \cite{Beloborodov_2013}. As we describe below, our first-principles plasma simulations demonstrate that most of the observed properties can indeed be naturally explained by coherent emission driven by a two-stream instability in a closed, twisted flux tube originating near the magnetic pole of the magnetar. 

\section{radiatively locked flow and two-stream instability}
\label{sec:theory}
\begin{figure*}[ht!]
\includegraphics[width=18cm]{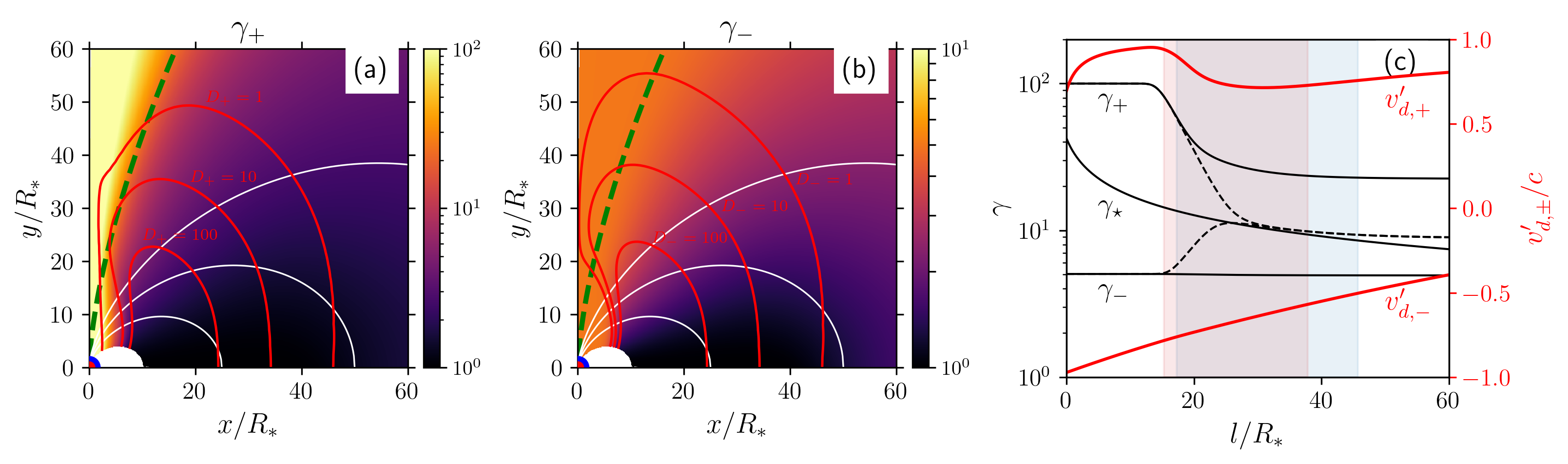}
\caption{Two-fluid model of the plasma flow in the magnetar magnetosphere. Panels (a) and (b) show Lorentz factors of positrons, $\gamma_+$, and electrons, $\gamma_-$. In this example, we have assumed that the $e^\pm$ outflow is injected at $r=2R_\star$ with a fixed multiplicity ${\cal M}=100$ (which remains constant along the outflow). The particles move
along the magnetic field lines (thick dashed green and white lines) and resonantly scatter thermal photons streaming from the surface of the magnetar. Radiation exerts a drag force on both electrons and positrons. Its local strength is characterized by the value of $D_\pm$ calculated in the model; the contours of $D_\pm=1,10,100$ are indicated by the red curves. In panel (c), we show $\gamma_\pm$ as well as $\gamma_\star$ along the thick dashed green field line, which has $R_{\rm max} = 900R_*$. The evolution of $\gamma_\pm$, when only the radiative force is considered, is shown with dashed lines. We also show velocities of electrons ($v_{d, -}'$) and positrons ($v_{d, +}'$) in the ``radiatively locked frame''. The orange and blue shaded regions have $D_+\geq3$ and $D_-\geq 3$, respectively.}
\label{fig:gamma}
\end{figure*}

The electric current density in an approximately dipole magnetosphere twisted through an angle $\psi\lesssim 1$ can be expressed as
\begin{equation}
    \boldsymbol{j}_0 = \frac{c}{4\pi} \frac{\boldsymbol{B}}{R_{\mathrm{max}}} \psi,
    \label{eq:current_density}
\end{equation}
where $R_{\mathrm{max}}$ is the maximum radius reached by the magnetic field line \citep{Beloborodov_2009}. This current is sustained by electron and positron motions along the magnetic field lines. Since magnetars rotate slowly, charge neutrality, $n_+ \approx n_- = n$, is a good approximation on the closed field lines (here, $n$ is the number density, 
and the subscripts ``$+$'' and ``$-$'' refer to positrons and electrons, respectively). Then, the existence of a non-zero current requires a difference between the velocities of the two species, $\delta v = |v_+ - v_-| \neq 0$. This difference must be sustained by an electric force against the radiative drag arising from resonant inverse Compton scattering of X-ray photons. Below we summarize a two-fluid model for the current in the presence of radiative drag that was constructed in \citet{Beloborodov_2013}.

An important parameter of the model is the pair multiplicity ${\cal M}$. It can be defined as ${\cal M} = (|j_+|+|j_-|)/|j_0| \approx (|v_+| + |v_-|)/\delta v$, where $j_{\pm}$ are the current densities carried by positrons and electrons, respectively. 
At high multiplicities ${\cal M}\gg 1$, the required current $j_0$ can be sustained by $e^+$ and $e^-$ outflowing from the star with nearly equal velocities, so that $\delta v\ll |v_\pm|$. In particular, for a relativistic outflow, one finds
$\delta v/c\sim 2/{\cal M}$. Plasma density can be further expressed as
\begin{equation}
    n = {\cal M} \frac{|j_0|}{e(|v_+| + |v_-|)} \approx \frac{\psi}{4\pi e} \frac{{\cal M} B }{(|\beta_+| + |\beta_-|) R_{\mathrm{max}}},
    \label{eq:number_density}
\end{equation}
where $e$ is the elementary charge, and $\beta_\pm = v_\pm/c$. In a steady state, $0 = \nabla \cdot \boldsymbol{j}_\pm = \nabla \cdot [(j_\pm/B)\boldsymbol{B}] = \boldsymbol{B} \cdot \nabla(j_\pm/B)$, which means that $j_\pm/B$ remains constant along the field lines, and thus so does $\cal{M}$. 

The radiative drag force $F_{\rm rad}$ continually
drives the particle velocities toward the ``radiatively locked velocity'' $v_\star$ at which $F_{\rm rad}$ vanishes (in the frame moving with $v_\star$, the radiation flux is perpendicular to the magnetic field lines and does not exert any force along $\bB$). A simple expression for $\beta_\star=v_\star/c$ is obtained by assuming a weakly twisted dipolar magnetosphere and a central radiation field (composed of photons streaming radially from the neutron star):
\begin{equation}
    \beta_\star (r, \theta) = \frac{B_r}{B} = \frac{2\cos\theta}{\sqrt{4\cos^2\theta + \sin^2\theta}},
\end{equation}
where $r, \theta$ are the spherical coordinates
and $B_r$ is the radial component of $\boldsymbol{B}$. 
As the drag force drives $v_+$ and $v_-$ toward the same value $v_*$, a parallel electric field $E_0$ must be generated to maintain the velocity separation $\delta v$ required by the electric current $j_0$.

The two-fluid dynamics of the $e^{\pm}$ flow is described by
\begin{equation}
    m_e c^2\frac{\mathrm{d}\gamma_\pm}{\mathrm{d}l} = F_{\rm rad} (\gamma_\pm) \pm e E_0,
    \label{eq:pair_equation_of_motion}
\end{equation}
where $m_e$ is the electron mass, $\gamma_\pm$ is the Lorentz factor of positrons/electrons and $l$ is the length along a magnetic
field line. Assuming thermal X-ray emission from the magnetar surface as the only source of radiation drag, the form of $F_{\rm{rad}}(\gamma)$ is derived in \cite{Beloborodov_2013},
\begin{equation}
    F_{\rm rad} = \frac{\alpha^2}{4} \frac{m_e c^2}{r_e} \left(\frac{R_*}{r}\right)^2 \left(\frac{k_B T_s}{m_e c^2}\right)^3 \gamma g(y) (\beta_\star - \beta),
\end{equation}
where $\alpha=e^2/\hbar c=1/137$, $r_e=e^2/m_ec^2$ is the classical electron radius, $T_s$ is the surface temperature, $k_B$ is the Boltzmann constant, $g(y)$ is the dimensionless Planck function evaluated at the resonant frequency $\omega_{\rm res}=\gamma^{-1}(1-\beta \cos\phi)^{-1}\omega_B$,
\begin{equation}
    g(y) \equiv \frac{y^3}{e^y - 1}, \quad y \equiv \frac{\hbar \omega_{\rm res}}{k_B T_s}.
\end{equation}
Here, $\phi$ denotes the angle between the photon propagation direction and the particle velocity, and $\omega_B$ is the cyclotron frequency. Combining Eq.~(\ref{eq:pair_equation_of_motion}) and the definition of multiplicity, $\cal{M}$, one finds
\begin{equation}
    m_e c^2 \frac{\mathrm{d}\gamma_+}{\mathrm{d}l} = \frac{F_{\rm rad}(\gamma_+) + F_{\rm rad}(\gamma_-)}{1 + \mathrm{d}\gamma_-/\mathrm{d}\gamma_+},
    \label{eq:positron_equation_of_motion}
\end{equation}
\begin{equation}
    \frac{\mathrm{d}\gamma_-}{\mathrm{d}\gamma_+} = \left(\frac{{\cal M} - 1}{{\cal M}+1}\right)^2 \left(\frac{\gamma_-}{\gamma_+}\right)^3.
\end{equation} 
It is easy to find $\gamma_\pm$ by integrating Eq.~(\ref{eq:positron_equation_of_motion}) along the field lines.

For example, suppose that the $e^\pm$ plasma is injected at radius
$r=2 R_*$ with fixed multiplicity ${\cal M}=100$ and high Lorentz factor $\gamma_+|_{l=0} = 100$. The initial Lorentz factor of electrons, $\gamma_-$, is determined by the relation $1-\beta_-/\beta_+ = 2/(\mathcal{M}+1)$, which follows directly from the definition of $\mathcal{M}$ under the assumption of charge neutrality. Assuming the magnetic field strength on the pole is $B_* = 3\times 10^{14}\, \mathrm{G}$, the computed $\gamma_+$ and $\gamma_-$ are shown in panels (a) and (b) of Figure~\ref{fig:gamma}. Only field lines with apex radii $R_{\rm{max}} \geq 10R_*$ are considered since plasma flows on field lines with small $R_{\max}$ likely have small multiplicities,
${\cal M} \sim 1$, and also tend to lose currents through the resistive untwisting process \citep{Beloborodov_2009}. One can see that the radiative drag force becomes strong at $r\gtrsim 10 R_*$, pushing $\gamma_\pm$ towards $\gamma_\star$. This process is further demonstrated in panel (c), where we plot $\gamma_\pm$ along a selected field line shown by the thick green dashed line, which has $R_{\rm max}=900R_*$. When only the radiative force is considered (i.e., neglecting $E_0$ in Eq.~\ref{eq:pair_equation_of_motion}), the evolution of $\gamma_\pm$ is shown with dashed lines. These clearly indicate that at $l \gtrsim 20 R_*$, without an electric field to sustain the velocity difference, the radiative force would quickly drive $\gamma_\pm$ toward $\gamma_\star$. The strength of the drag can be quantified using the dimensionless parameter,
\begin{equation}
    D_\pm \equiv \bigg|\frac{r F_{\rm{rad}}(p_\pm)}{c(p_\pm - p_\star)}\bigg|,
\end{equation}
where $p$ is the momenta of the particles, $p_\star\equiv \gamma_\star m_e v_\star$, and $\gamma_\star \equiv 1/\sqrt{1 - \beta_\star^2}$. Parameter $D$ represents the ratio of the local light crossing time, $r/c$, to the cooling time, $t_{\rm{cool}} \equiv -(p-p_\star)/F_{\rm{rad}}$.\footnote{Note that the definition of $D$ used here differs from that in \cite{Beloborodov_2013}; consequently, the $D$ contours shown here are also different.} 
The contours of $D_\pm$ shown in panels (a) and (b) indicate that cooling is dynamically strong, $D_\pm \gtrsim 1$, in the region $10\lesssim r/R_* \lesssim 50$. In the polar field line zone where we find bright radio emission (see Section~\ref{sec:global_emission}), $1\lesssim D \lesssim 100$. 

When the plasma flow enters the strong cooling region, the drag force pushes the velocities of both species toward $v_\star$, leading to $\delta v \ll v_\pm \approx v_\star$. Therefore, it is convenient to study the dynamics in the ``radiatively locked frame'' (RL frame) that moves along the magnetic
field lines with velocity $v_\star$. In this frame,
the drag force may be approximated by a simple form,
\begin{equation}
    F'_{\rm rad} \approx -p'/t'_{\rm cool},
    \label{eq:drag_rlFrame}
\end{equation}
where $t'_{\rm cool} = t_{\rm cool}/\gamma_\star$, and prime
indicates that the quantity is measured in the RL frame.
In the RL frame, electrons and positrons move approximately at the same speed, $v'_d$, but in the opposite direction (see panel (c) in Figure~\ref{fig:gamma}). When $\gamma_\star^2 / {\cal M} < 1$, one has 
\begin{equation}
    v'_d/c \sim \gamma_\star^2 \delta v / (2c) \sim \gamma_\star^2 / {\cal M}.
    \label{eq:v_dr}
\end{equation}

As the electric field $E_0 \sim F_{\rm rad}(v'_d)/e$ maintains the velocity separation, the combination of the electric and radiation forces squeezes the distribution functions of $e^{\pm}$ toward $\delta$-functions centered at $\pm v'_d$. The two-stream instability (which is not taken into account in the two-fluid model) is inevitably triggered when the thermal spread of a single beam becomes comparable to the separation between the two beams\footnote{For non-relativistic Lorentzian distributions, this result can be shown exactly, see Appendix~\ref{appen:stability_threshold} for more details.}, $p'_d = m_e v'_d/\sqrt{1 - (v'_d/c)^2}$. Therefore, in a full kinetic model, the continual radiative drag operating on the timescale $t_{\rm cool}$ fails to establish the two cold $e^\pm$ streams with velocities $\pm v_{d}'$. Instead, it leads to sustained turbulence. Indeed, for the conditions in the zone of active two-stream instability, $t_{\rm cool}$ is much longer than the plasma timescale (which determines the instability growth rate):
\begin{equation}
    \omega'_p t'_{\rm cool} \approx 10^4 \bigg(\frac{\omega'_p}{10^9\mathrm{rad/s}}\bigg)\bigg(\frac{r}{30R_*}\bigg)\bigg(\frac{10}{D}\bigg)\bigg(\frac{10}{\gamma_\star}\bigg),
    \label{eq:cooling}
\end{equation}
where $\omega'_p \equiv \sqrt{4\pi n' e^2/m_e}$ is the local cold plasma frequency in the RL frame, and $R_* = 10\,\mathrm{km}$.

The sustained turbulence can continually generate low-frequency electromagnetic radiation over a range of distances where the cooling is strong, $D\gtrsim 1$. This emission process is demonstrated by direct kinetic simulations below.

\section{Simulation Method and Setup}
\label{sec:setup}

As discussed in Section~\ref{sec:theory}, the cooling time is much longer than the plasma period, which leads to both a small power of the coherent emission and slow growth of the cooling-induced instability. These conditions present a challenge for the standard particle-in-cell (PIC) plasma simulation technique, as these weak signals can be overwhelmed by Poisson noise. Therefore, we decided to simulate the cooling-driven instability using a relativistic extension of the continuum code Gkeyll \citep{Juno_2018,HakimJuno_2020}, which solves the Vlasov-Maxwell system of equations using a discontinuous Galerkin scheme (Juno et al., in preparation). Results of our simulations are numerically converged for all aspects of the analysis. Simulations are performed in the RL frame. In Section~\ref{sec:theory}, quantities measured in this frame were denoted with a prime; however, in Sections~\ref{sec:setup}-\ref{sec:2D} below we will omit the prime to simplify the notation, as we do not use any other frames in these sections.

In the inner magnetar magnetosphere, fast synchrotron losses confine  particle motions strictly along the magnetic field.\footnote{Transverse particle motions can exist in the outer magnetosphere, $r\gtrsim 10^3R_*$, and one-dimensional electromagnetic streaming instabilities with transverse motions are possible in this region. However, they do not lead to plausible radio emission mechanisms, as shown in Appendix~\ref{app:transverse_instab}.} Therefore, we solve the plasma kinetic equation in one-dimensional velocity space (1V):
\begin{equation}
    \frac{\partial f}{\partial t} +  \frac{\partial}{\partial x}\left(\frac{p_x}{m_e\gamma}f\right) + \frac{\partial}{\partial p_x}\bigg(f(qE_x + F_{\rm rad})\bigg) = 0,
\end{equation}
where the $x$-axis is chosen along the background magnetic field $\boldsymbol{B}_0$. We decompose the full magnetic field, 
$\boldsymbol{B}_{\rm full}$, into a sum of the background field $\bB_0$, the component describing magnetospheric twist, $\boldsymbol{B}_{\rm  tw}$, and the field perturbation $\bB$
generated during the instability.  Ampere's law, then, may be written as
\begin{equation}
\label{eq:Ampere}
    c \nabla \times \boldsymbol{B} = 4\pi (j - j_{0})\hat{x} + \frac{\partial \boldsymbol{E}}{\partial t},
\end{equation}
where $c(\nabla \times \boldsymbol{B}_{\rm  tw})/4\pi=j_0 \hat{x}=-2n_e ev_d$. One-dimensional simulations are electrostatic and do not include the field perturbation
$\boldsymbol{B}$, which corresponds to setting $\bB=0$ in Eq.~(\ref{eq:Ampere}). Two-dimensional simulations track the evolving $\bB$ using the Faraday's induction law, $\partial \boldsymbol{B}/{\partial t}=-c\nabla \times \boldsymbol{E}$.

\begin{figure*}[ht!]
\includegraphics[width=18cm]{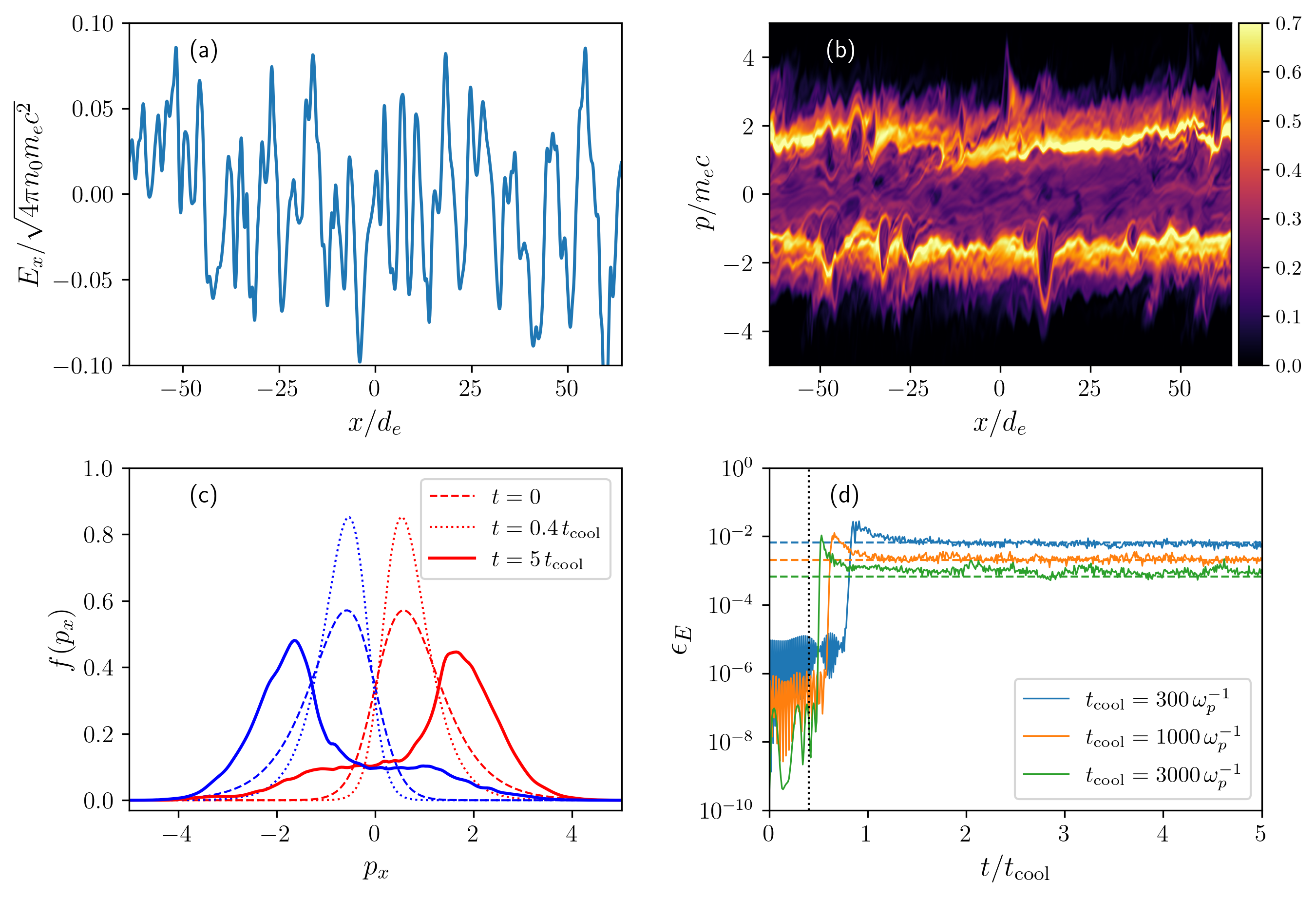}
\caption{Results of the 1X1V simulations. Panels (a) and (b) show snapshots of the electric field and distribution functions in the phase space, $f(x,p_x)$, at the end of the simulation with $v_d/c=0.5$ and $\omega_p t_{\rm cool}=3000$, measured at $t = 5t_{\rm cool}$. Panel (c) shows the distribution functions, $f(p_x)$ at the beginning of the simulation, just before the onset of the instability, and at the saturation. Panel (d) shows the evolution of electric field energy density normalized to kinetic energy density, $\epsilon_E$, for different cooling times, $t_{\rm cool}$. The vertical dotted line marks $t=0.4 t_{\rm cool}$, which corresponds to the moment just before the onset of the instability. The saturated turbulence level is consistent with $\epsilon_E\approx 2(\omega_p t_{\rm cool})^{-1}$ (indicated by the horizontal dashed lines) for all three simulations with  different $t_{\rm cool}$.}
\label{fig:1x1v_gkyl}
\end{figure*}

The initial state of the electrons and positrons is set up with the Maxwell-Juttner (MJ) distribution,
\begin{equation}
    f_{\pm} = \frac{n_0}{2K_1(\xi)\gamma_d}\exp[-\gamma_d\xi (\gamma \pm v_dp)],
\end{equation}
following the procedure described in \cite{Johnson_2025}. Here, $\gamma_d \equiv 1/\sqrt{1 - \beta_d^2}$, $K_1$ is the Macdonald function, $\xi \equiv m_ec^2/k_BT$, and $T$ is the plasma temperature.

Plasma in the simulation box is subject to the drag force of
the form given in Eq.~(\ref{eq:drag_rlFrame}). Hereafter, the radiative drag is also called ``cooling,'' since it tends to shrink the distribution function toward $p=0$ in the RL frame. The cooling strength is set uniform across the simulation domain. To keep the timespan of simulations reasonable, we choose the cooling time so that $\omega_p t_{\rm cool} = 300, 1000, 3000$. While these values are smaller than expected in magnetars (see Eq.~\ref{eq:cooling}), the results of our simulations can be reliably extrapolated to realistic conditions.

Below we describe the simulation results in one-dimensional (1X1V) and two-dimensional (2X1V) configuration spaces in Sections~\ref{sec:1D} and~\ref{sec:2D}. In two-dimensional simulations, electric and magnetic fields are allowed to vary across field lines, which enables excitation of electromagnetic modes.

In 1X1V, we focus on the simulations with $v_d/c = 0.5$, $k_BT/m_e c^2 = 0.3$, and no initial perturbation in the electric field. The extent of the configuration space is $[-L_x/2, L_x/2] = [-64d_e, 64d_e]$, where $d_e \equiv c/\omega_p=\sqrt{m_e c^2/4\pi n_0 e^2}$ is the cold skin depth, and the extent of the momentum space is $[-64m_ec, 64m_ec]$. The configuration space is resolved by $N_x = 512$ cells, and the momentum space is resolved by $NV_x = 1024$ cells.\footnote{Gkeyll's newest implementation, not yet tested at the time we carried out the simulations reported in this Letter, enables the use of non-uniform momentum-space grids and higher-order third-degree polynomial basis functions, thereby reducing the number of grid points needed and lowering computational costs.} We employ piecewise quadratic Serendipity elements for the discontinuous Galerkin basis expansion \citep{Arnold_2011}. Periodic boundary conditions are applied in the configuration space, while zero-flux boundary conditions are used in the momentum space. We have also performed simulations with $v_d/c = 0.25, 0.8$, varied initial temperatures, and a strong initial electric field perturbation.

In 2X1V, our numerical setup allows the escape of waves in the $ y$-direction, which is transverse to the magnetic field lines. The current-carrying zone (z3) has a finite $y-$extent. Outside this zone, we first gradually decrease $v_d$ and $j_{0}$ to zero (which we refer to as current transition zones, z2), and then reduce the plasma density (forming density transition zones, z1). The density is decreased to zero near the $y$-boundaries, so that the plasma-filled region of the computational box is separated from the boundaries by vacuum zones (z0). Furthermore, near the $y$-boundaries, we introduce an absorbing layer to damp escaping electromagnetic waves. This setup enables waves generated in the central current zone to propagate towards the vacuum zones and leave the plasma. We use the Poynting flux of waves escaping through the absorbing layers as a proxy for the luminosity of the low-frequency electromagnetic emission generated by turbulent plasma. 

The two-dimensional simulations are performed with three drift speeds, $v_d/c = 0.25, 0.5, 0.8$. In all three cases, we set the initial plasma temperature $k_B T/m_ec^2 =2 (v_d/c)^2$, so that the system is initially stable. The initial state has no electric field and no perturbation of the magnetic field. The simulation domain has a physical size of $L_x \times L_y = 50d_e\times 160d_e$. For $v_d/c=0.5$ and $0.8$, the domain is resolved by $N_x \times N_y = 400 \times 1280$ cells, while for $v_d/c=0.25$, a finer resolution of $800 \times 2560$ is used. The momentum space ranges are $[-32m_e c, 32m_e c], [-48m_e c, 48m_e c],$ and $[-64m_e c, 64m_e c]$ for $v_d/c=0.25, 0.5$ and $0.8$, respectively, and are resolved by $NV_x= 2048, 1536, 2048$ cells.

\section{1D Simulations: trigger and saturation of two-stream instability} \label{sec:1D}

The results from a set of simulations with $v_d/c=0.5$ are summarized in Figure~\ref{fig:1x1v_gkyl}. Most of the presented runs used the cooling strength parameter $\omega_p t_{\rm cool} = 3000$; however, we have found similar results for different cooling strengths. Immediately after the simulation begins, an electric field develops to counteract the radiation drag and sustain the required current $j_0$. As expected, the magnitude of this field is $E_0 \approx F_{\rm rad}(\gamma_d)/e$.

The combined electric and radiative forces then drive the system toward the state predicted by the two-fluid model (Section~\ref{sec:theory}), squeezing the $e^\pm$ distributions toward two delta-functions on the timescale $\sim t_{\rm cool}$. When the two distributions become narrow enough to develop a large gap between them, the two-stream instability is triggered. This happens when the thermal spread in both $e^+$ and $e^-$ beams becomes comparable to the velocity separation between the two beams, which occurs at $t\approx 0.4 t_{\rm cool}$ (see panel 
(c) of Figure~\ref{fig:1x1v_gkyl}). The instability exponentially amplifies the electric field oscillations on a timescale shorter than the cooling time. The fast exponential growth of the energy density of the electric field  is observed until the turbulence level approaches saturation (panel d).

The instability saturates when the amplified electrostatic waves trap the initially counter-streaming $e^{\pm}$ beams. The trapping leads to a violent redistribution of the particle momenta and the formation of a flat plateau in the distribution function,
$\partial f/\partial p_x\sim 0$, at small momenta $p_x$ in the middle between the $e^\pm$ peaks. The plateau is composed of relatively slow particles, $|v|<v_d$, and it develops together with an overall broadening of the distribution functions $f_\pm(p_x)$, so that the average $e^\pm$ velocities remain large enough to conduct the required current $j_0$. As a result, the peaks of the $e^\pm$ and distributions shift to $\pm v_{\rm peak}$. For sufficiently weak cooling ($\omega_p t_{\rm cool} = 3000$ in our simulations), the ratio $p_{\rm peak}/p_d$ shows a weak dependence on $v_d$, where $p_{\rm peak}$ is the momentum corresponding to $v_{\rm peak}$. We measured $p_{\rm peak}/p_d \sim 2.5$ for $v_d/c$ ranging from 0.25 to 0.8. 

The simulation demonstrates the formation of a quasi-steady state, with the plateau level self-regulated to maintain the distribution function near marginal instability, resulting in a significantly reduced growth rate. A small growth rate of the instability, $\Gamma\sim t_{\rm cool}^{-1}\ll\omega_p$ is sufficient to sustain the electrostatic turbulence that kicks the particles and offsets the cooling effect of the radiative drag.

This saturated state is characterized by a strong stochastic electric field, $E_x$ (panel a). Despite the large amplitude of $E_x$, its average over space and time, $\langle E_x \rangle$, remains comparable to $E_0$. Furthermore, the observed time-averaged rate of ohmic dissipation is consistent with effective $E_x\sim E_0$: $\langle \textbf{j}\cdot \textbf{E}_x\rangle\sim j_0 E_0$.\footnote{Rephrasing the same result, the dissipation due to
turbulent resistivity, $\langle \delta\boldsymbol{j}\cdot \delta\boldsymbol{E}\rangle$, may be neglected compared to the dissipation driven by the average electric field $E_0$.} 
Thus, the large turbulent voltage fluctuations do not affect the net dissipation rate and the corresponding lifetime of the magnetospheric twist, in agreement with \cite{Beloborodov_2013}. 

We also find that the ratio of the electrostatic energy density $E^2/8\pi$ to the plasma kinetic energy ${\cal E}_K=(\langle\gamma\rangle-1)n_0 m_e c^2$ is small in the steady state,
$\epsilon_E \equiv E^2/(8\pi {\cal E}_K) \ll 1$. Nevertheless, the sustained electrostatic turbulence is not weak, as it remains marginally capable of trapping particles. This regime approximately corresponds to the so-called ``critical balance'' regime of electrostatic turbulence \citep{Ewart_2025, Nastac_2025}, where the timescale of the plasma oscillation in the electrostatic wave is comparable to the timescale for nonlinear effects. Using the corresponding terms of the kinetic equation, one can express this condition as $k\Delta v\sim eE/(m_e\Delta v)$, where $\Delta v$ is the saturated spread of the distribution function. Particle trapping is clearly observed in phase space, e.g., in the snapshot shown in Figure~\ref{fig:1x1v_gkyl} (panel~b). It is seen as coherent holes with a spatial scale $\sim 5d_e$. The appearance of coherent structures is ubiquitous in simulations for all cooling times we consider. 

The size of the coherent structures approximately corresponds to the wavelength $\lambda_m$ of the peak of the sustained turbulence spectrum. We find from our simulations that $\lambda_m$
scales as $\sim \sqrt{t_{\rm{cool}}}$.\footnote{This scaling can be explained by considering the scaling of the growth rate, $\Gamma$, near the marginal stability limit, $\Gamma\propto k^2$, and the condition that in steady state the growth time of the instability should be roughly equal to the cooling time, $1/\Gamma\sim t_{\rm cool}$ (Popova, Zeng et al., in prep).} The larger $\lambda_m$ at longer $t_{\rm cool}$ implies that a smaller wave amplitude becomes sufficient to trap the particles: $ E_x\propto\lambda_m^{-1}\propto t_{\rm cool}^{-1/2}$. The condition for trapping, $eE_x \lambda_m \sim (\gamma_d - 1)m_e c^2$, results in $\epsilon_E \sim (\omega_pt_{{\rm cool}})^{-1}$, which matches well with the measured scaling shown in panel~(d) of Figure~\ref{fig:1x1v_gkyl}.\footnote{A more simplistic quasi-linear argument provides the same scaling, $\epsilon_E \sim (\omega_p t_{\rm{cool}})^{-1}$ \textemdash  see Appendix~\ref{appen:scaling_quasi_linear} for more details.}

\begin{figure*}[ht!]
\includegraphics[width=18cm]{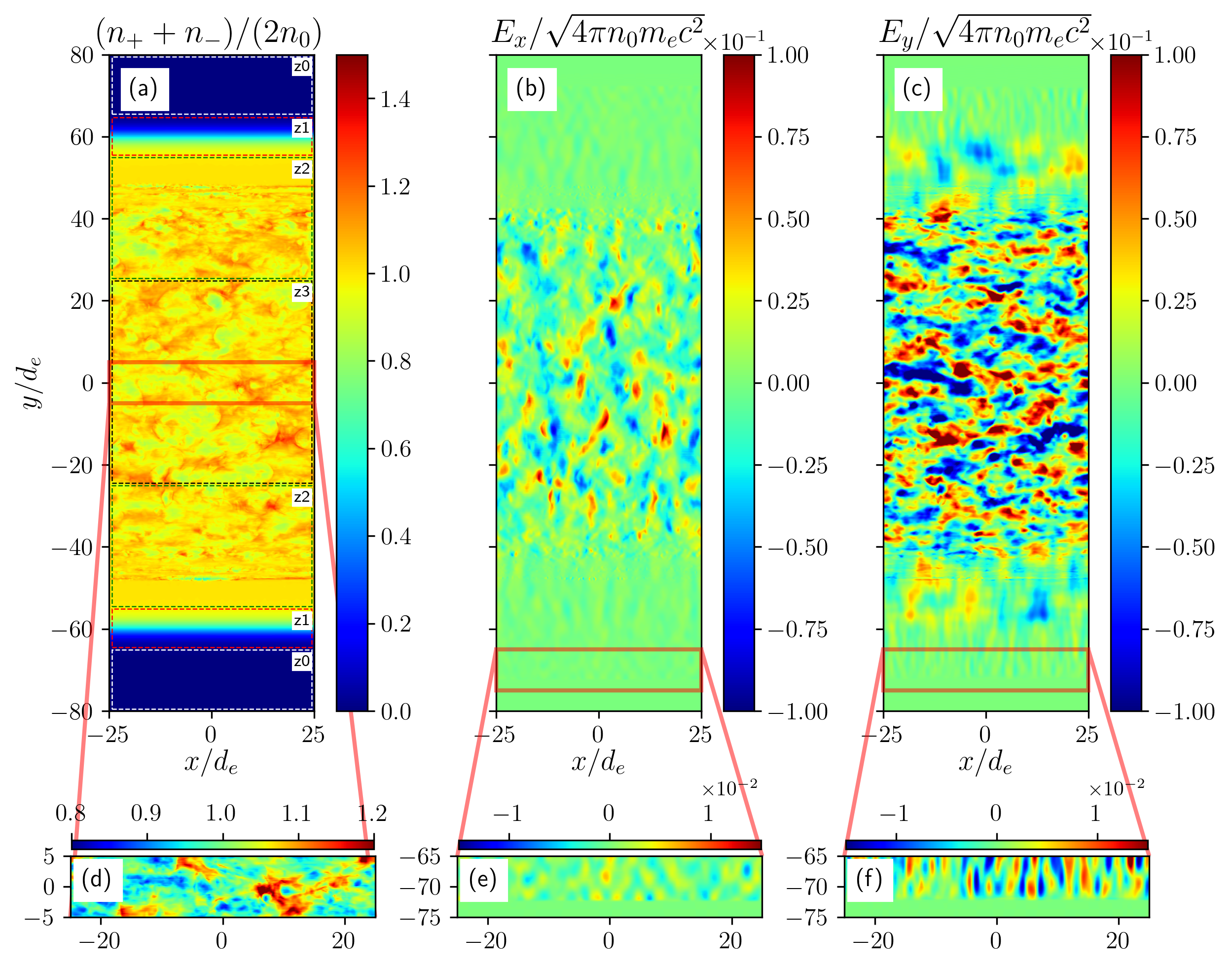}
\caption{Snapshot during the quasi-steady state of the 2X1V simulation for $v_d/c=0.5$, $\omega_pt_{\rm cool}=3000$. Panels (a)-(c) show the snapshots of plasma density, $n_{+}+n_{-}$, and electric field components, $E_x$, $E_y$, displaying the excitation of turbulence in the current-carrying zone, as well as the production and propagation of electromagnetic waves. In panel (a), different zones are indicated with dashed boxes and labeled as follows: z0 is the vacuum zone, z1 is the density transition zone, z2 is the current transition zone, and z3 is the current zone. The inset (d) shows density fluctuations in the current-carrying zone, and insets (e)-(f) show electromagnetic waves escaping through the absorbing layers. 
\label{fig:2D_gkyl}}
\end{figure*}

The steady state achieved in our simulations is independent of the initial conditions. After initial relaxation, all the simulations with varied initial temperatures and strong initial turbulent electric fields converge to the same steady state of particle distribution functions, turbulent field energy density, and spectrum. 

\section{2D simulations: excitation of electromagnetic modes} \label{sec:2D}

In this Section, we extend our simulations to a 2D configuration space, 2X1V. Now perturbations can vary in $x$ and $y$, i.e. their wavevectors $\boldsymbol{k}$ can have any direction in the $x$-$y$ plane. This enables the generation of escaping electromagnetic waves, as described below.

\begin{figure*}[ht!]
\includegraphics[width=18cm]{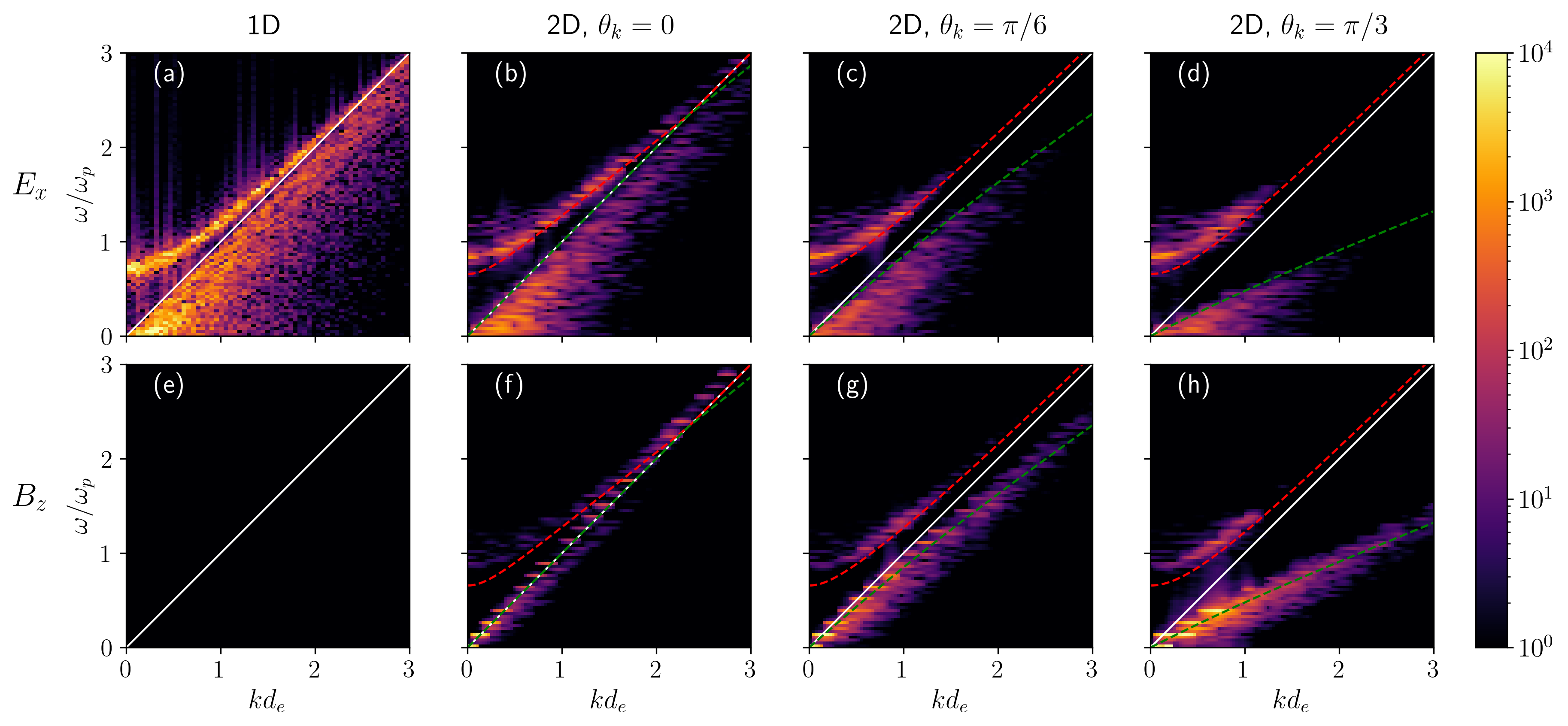}
\caption{Dispersion of the plasma modes in 2X1V simulation for $v_d/c=0.5$, $\omega_pt_{\rm cool}=3000$. Intensities in the Fourier space, $(\omega-k)$, are shown for electric, $E_x$, panels (b)-(d), and magnetic, $B_z$, field components, (f)-(h), for different angles of propagation, $\theta_k=0, \pi/6, \pi/3$. The red and green dashed lines correspond to the dispersion relations of the superluminal and Alfven modes, respectively. The white lines mark the vacuum light wave dispersion, $\omega = kc$. For reference, panels (a) and (b) show corresponding Fourier intensities in the 1X1V simulation with same parameters.
\label{fig:omega_k}}
\end{figure*}
Similar to the 1D simulations, a net electric field $E_0 \approx F_{\rm rad}/e$ immediately develops in the central current zone, and the two-stream instability is triggered when the counter-streaming beams become sufficiently narrow in momentum space. In Appendix~\ref{appen:eigenmodes}, we calculate the growth rate and polarization of linearly unstable modes in the idealized case of cold beams. We find that the two-stream instability pumps non-propagating purely growing, $\rm{Re}(\omega)=0$, electromagnetic perturbations in the ordinary polarization mode.\footnote{The extraordinary X-modes  (which have $\boldsymbol{E}$ oscillating along $z$) do not grow during the linear or non-linear phases of the instability. They are fully decoupled because the magnetized $e^\pm$ plasma does not support electric currents in the transverse $y$-$z$ plane.} This mode has the electric field in the plane containing the wavevector $\boldsymbol{k}$ and the background magnetic field, so the perturbation field components are $E_x$, $E_y$, and $B_z$. At sufficiently low wavenumbers, $k^2 < 2\omega_p^2/(\gamma_d^3 v_d^2)$, unstable modes exist for any direction of $\boldsymbol{k}$. The instability has the growth rate $\Gamma\propto \cos \theta_k$, where $\theta_k=\arctan(k_y/k_x)$ is the wave propagation angle relative to the background magnetic field. The fastest-growing modes are the electrostatic waves aligned with the magnetic field 
($\theta_k=0$), same as in the 1D simulations.

The exponential growth of the unstable modes ends when they reach an amplitude capable of trapping the initial beams. At that point, the amplified nonpropagating modes break, leading to the nonlinear
electromagnetic turbulence of waves with various directions of $\boldsymbol{k}$. After a relaxation period of roughly $1.5t_{\rm cool}$ following the onset of the instability, the system settles into a steady state, where the sustained turbulence in the current-carrying zone continually sources electromagnetic fluctuations. A fraction of these modes propagate toward the vacuum zone and escape through the absorbing layer near the $y$-boundaries. This behavior is illustrated in the snapshot of the simulation shown in Figure~\ref{fig:2D_gkyl}, with $v_d/c=0.5$ and $t_{\rm cool}=3000\omega^{-1}_p$.

To study the normal modes generated by the turbulence, we have calculated the temporal and spatial Fourier transforms of the fields $ E_x$, $ E_y$, and $ B_z$ that represent ordinary polarization. Their Fourier amplitudes are expected to peak on the dispersion curve $\omega(\boldsymbol{k})$ of the normal modes. Figure~\ref{fig:omega_k} shows the intensities of Fourier components of $E_x$ and $B_z$, $|E_{x}(\omega,{\textbf{k}})|^2,|B_{z}(\omega,{\textbf{k}})|^2$,  on the plane of $\omega$ and $k=\sqrt{k_x^2+k_y^2}$ for a fixed propagation angle $\theta_k$ (different panels correspond to $\theta_k=0,\pi/6,\pi/3$). For comparison, we also show the 1D simulation results. One can see that the perturbations of $E_x$ and $B_z$ do peak along the expected dispersion relation for the two well-known branches of the ordinary mode \citep{Arons_1986}: superluminal and subluminal (Alfv\'en), here
computed for the cold $e^\pm$ beams with momenta $\pm p_{\rm peak}$. The Fourier spectrum of $E_x$ also shows a third branch at lower frequencies. Excitation of this mode is strongest for the
field-aligned propagation, $\theta_k\sim 0$; it is also present in 
the 1X1V simulations. We associate these fluctuations with additional electrostatic modes sourced by the plateau of the distribution function, where $\partial f/\partial p_x\sim 0$, similar to ``electron acoustic waves'' \citep[e.g.,][]{Valentini_2012}.\footnote{In particular, we have reproduced the existence of these modes by calculating linear dispersion characteristics of 1D plasma containing two beams and a plateau (Popova, Zeng et al., in preparation).}

Similar to the problem of radio emission in pulsars \citep{Philippov_2020}, the superluminal mode is of special interest since, unlike the Alfv\'en mode, it does not suffer from Landau damping and can escape through a plasma of decreasing density $n$. As the superluminal mode travels through the plasma with decreasing $\omega_{\rm p}\propto \sqrt{n}$, it gradually becomes a usual electromagnetic wave with propagation speed approaching $c$ when $\omega_{\rm p}\ll\omega$. In our simulations, these modes propagate through the plasma and escape through the absorbing layers near the $y-$boundaries (see also Appendix~\ref{appen:superluminal_propagation}). At low $k$, the superluminal dispersion curve starts at the relativistic plasma frequency $\omega_{\rm min} = \sqrt{2} \omega_{\rm p}/\gamma_{\rm peak}^{3/2}$, where $\gamma_{\rm peak} = 1/\sqrt{1 - v_{\rm peak}^2}$. Its energy density observed inside the turbulence excitation region is concentrated around the local $\omega_{\rm min}$ (see Fig.~\ref{fig:omega_k}). As a result, the characteristic frequency of escaping radiation is close to $\omega_{\rm min}$ defined inside the current-carrying region of the simulation box:
$\omega_{\rm em}\approx\omega_{\rm min}$. 

\begin{figure}[ht!]
\includegraphics[width=\linewidth]{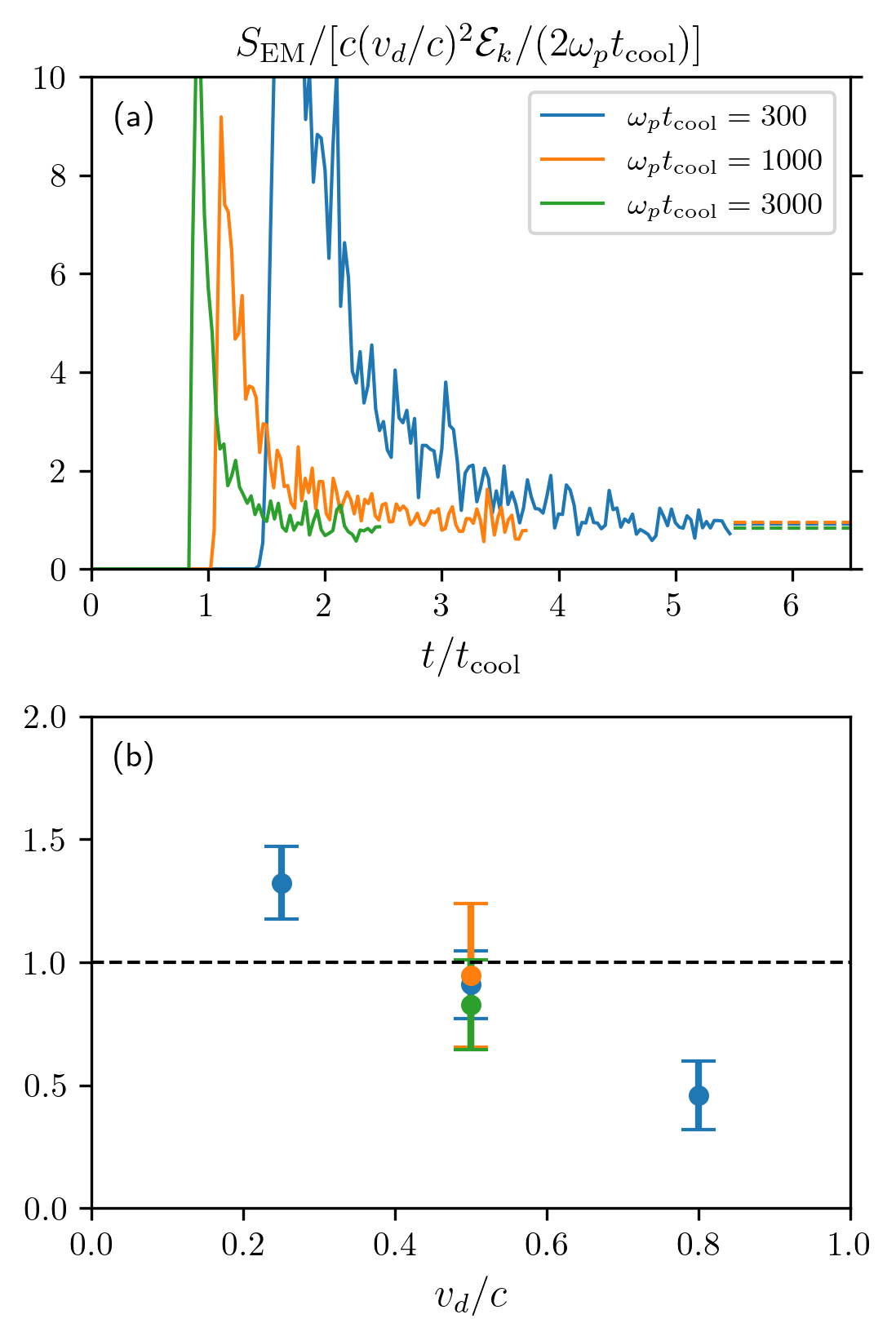}
\caption{Dependence of the normalized Poynting vector, $S_{\rm EM}$, of electromagnetic radiation escaping through the $y$-boundaries of 2X1V simulations on the cooling strength (panel (a), for a fixed $v_d/c=0.5$),  and the drift speed (panel b). Blue, orange and green data points in both panels correspond to simulations with cooling times $\omega_p t_{\rm cool}=300,1000,3000$. The values represented by the dashed lines in panel (a) and the points in panel (b) are calculated from the data taken over the last $0.5 t_{\rm cool}$ of each simulation. Errorbars represent fluctuations of the measured Poynting flux. 
\label{fig:poynting_flux}}
\end{figure}

The energy flux of escaping electromagnetic waves $S_{\rm EM}$ is described by the $y$-component of the Poynting vector $S_y =-c E_x\times B_z/4\pi$ measured in the boundary vacuum zones.\footnote{In simulations conducted in periodic boxes, we find that the energy flux of superluminal waves is essentially isotropic in the RL frame. Consequently, we expect $S_y$ to be representative of the flux in any direction.
} We define $S_{\rm EM}\equiv(\int_x S_y {\rm{d}} x)/L_x$ by averaging $S_y$ along the $y$-boundaries of the simulation box. We find that for sufficiently large boxes $S_{\rm EM}$ shows only weak dependence on the box size. This indicates a saturated value for
the energy density of superluminal waves, ${\cal E}_{\rm EM}\approx S_{\rm EM}/c$. We interpret this saturation as a balance established by the nonlinear interactions of plasma modes, partitioning the turbulence energy ${\cal E}_{\rm turb}$ between different modes, where ${\cal E}_{\rm turb}$ is a fraction of the total kinetic energy density of the plasma, ${\cal E}_K$. The energy density of superluminal waves then may be expected to scale with the total energy density of electrostatic waves obtained in 1D simulations, $\sim {\cal E}_K/(\omega_p t_{\rm cool})$.

To study how the radiated power depends on the parameters of the problem, we performed two sets of simulations: (1) we fixed the drift velocity $v_d=0.5c$ and varied $t_{\rm cool}$ from $300\omega_p^{-1}$ to $3000\omega_p^{-1}$, and (2) we fixed $t_{\rm cool}=300\omega_p^{-1}$ and varied $v_d$ from $0.25c$ to $0.8c$. The results are well approximated by the simple analytical formula,
\begin{equation}
    \frac{S_{\rm EM}}{c} \approx \frac{1}{2}\left(\frac{v_d}{c}\right)^2\left(\frac{{\cal E}_K}{\omega_p t_{\rm cool}}\right),
    \label{eq:power_RL}
\end{equation}
where ${\cal E}_K/t_{\rm cool} \sim j_0 E_0$ describes the ohmic dissipation power resulting from the radiation drag.

The time evolution of $S_{\rm EM}$ in simulations with $v_d=0.5 c$ and different $t_{\rm cool}$ is shown in panel (a) of Figure~\ref{fig:poynting_flux}. In all cases, $S_{\rm EM}$ normalized to the value given by Eq.~(\ref{eq:power_RL}) greatly exceeds unity during the initial evolution phase while the instability saturates. This initial burst is associated with the conversion of the exponentially growing, non-propagating modes into a continuum of propagating waves, some of which are superluminal modes. Once the steady state is approached, i.e., roughly at $t>2t_{\rm cool}$, the sustained turbulence continues to produce $S_{\rm EM}$ at a level consistent with Eq.~(\ref{eq:power_RL}).

For our $v_d$ parameter scan, we also find the approximate consistency of the steady-state $S_{\rm EM}$ with Eq.~(\ref{eq:power_RL}) (panel (b) of Figure~\ref{fig:poynting_flux}). The nonlinear saturation has a strong limiting effect on $S_{\rm EM}$ at $v_d\gtrsim 0.5$. In the model with $v_d=0.8c$, the observed $S_{\rm EM}$ falls below the analytical expression that uses ${\cal E}_K$ in the midplane $y=0$. This model has so strong turbulence that the zone of nonlinear saturation of the superluminal wave density ${\cal E}_{\rm EM}$ extends from the midplane to larger $y$ where our setup has a weaker electric current and hence a lower ${\cal E}_K$. The observed $S_{\rm EM}$ may become consistent with Eq.~(\ref{eq:power_RL}) when the appropriately reduced ${\cal E}_K$ is used in the analytical expression. We leave further study of this effect for future work. 

For all simulations presented in this work, we also performed parallel simulations of the same setup using the PIC code Tristan-v2 \citep{hayk_hakobyan_2023}. We were unable to converge the results of PIC simulations for the amplitude of the weak initial electric field, $E_0$, and the peak amplitude of the coherent emission for our longest cooling time simulation, $t_{\rm cool}=3000\omega^{-1}_p$. Despite this limitation, PIC simulations gave converged results for the steady-state emission rate, which closely matches with our Gkeyll simulations.

\section{A global model of radio emission} \label{sec:global_emission}

\begin{figure}[ht!]
\includegraphics[width=\linewidth]{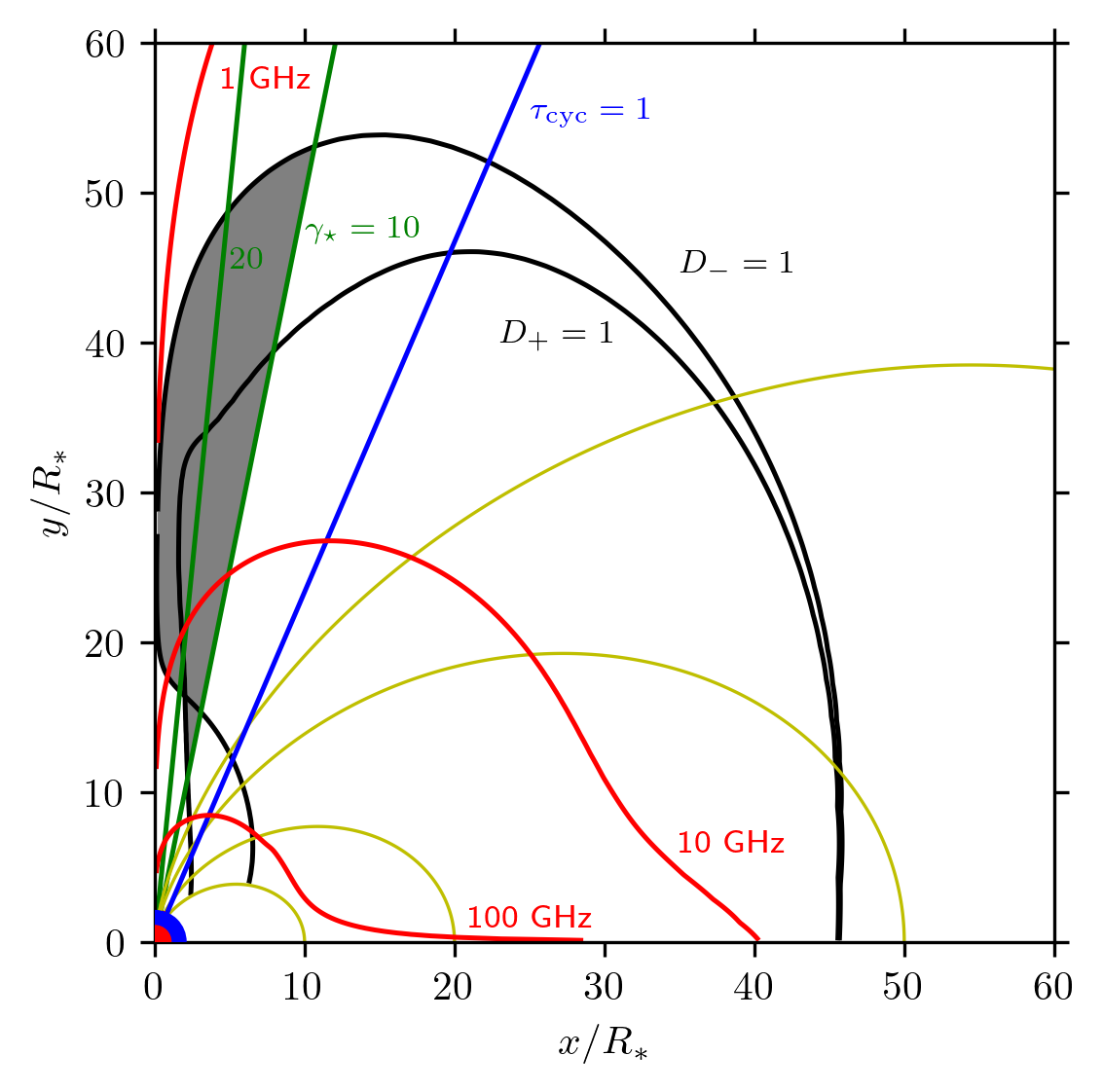}
\caption{Global model of radio emission from the magnetar magnetosphere. The radio-emitting zone is shaded in gray. Electrons and positrons experience a strong radiative drag,
$D_{\pm}>1$, in the region bound by black contours. Green curves highlight contours of $\gamma_\star=10$ and 20, sufficient for production of powerful radio emission. Red curves are the contours of emission frequency $\nu_{\rm em}$, which we take comparable to the Doppler-boosted plasma frequency, $\gamma^{1/2}_*\nu_p$. The blue curve shows the boundary of the region that can produce radiation capable of escaping through the outer magnetosphere (the optical depth to cyclotron absorption remains below unity along the escape path). In this calculation, we adopted ${\cal M} = 50$.} 
\label{fig:power_freq}
\end{figure}

The found expressions for the emission power and frequency in the local RL frame allow one to evaluate the global picture of radio emission from the twisted magnetosphere. This requires transformation of the emitted radiation from the RL frame to the global static lab frame (the rest frame of the magnetar). We now return to notation in Section~\ref{sec:theory} where all quantities measured in the RL frame are denoted with a prime to distinguish them from quantities measured in the lab frame.

In the RL frame, the emission has the characteristic frequency $\nu'_{\rm em}\approx {\sqrt{2}\nu_p'}/{\gamma_{\rm peak}^{3/2}}$ (section~\ref{sec:2D}), where $\nu'_p = \omega'_p/2\pi$. Emission at an angle $\delta'$ relative to $\bB$ is Doppler-boosted to the lab frame by the factor ${\cal D}=\gamma_\star(1+\beta_\star\cos\delta')$. The average angle $\delta'=\pi/2$ defines the average Doppler factor ${\cal D}=\gamma_\star$ and a characteristic beaming angle $\delta$ of emission in the lab frame: $\sin\delta=\gamma_{\star}^{-1}$. When $\gamma_\star\gg 1$, the  emission is strongly beamed along the magnetic field lines. The maximum Doppler-boosted frequency occurs at $\delta'=0$:
\begin{equation}
    \nu_{\rm em} = \gamma_*({1 + \beta_\star}) \nu'_{\rm em} \approx \frac{\sqrt{2} (1 + \beta_\star) \gamma_\star^{1/2}}{\gamma_{\rm peak}^{3/2}} \nu_p,
    \label{eq:freq}
\end{equation}
where $\nu_p=\sqrt{4\pi n e^2/m_e}/2\pi =\gamma_\star^{1/2} \nu_p'$ (since $n=\gamma_\star n'$), and the plasma density $n$ is given by Eq.~(\ref{eq:number_density}).
As shown below, emission should occur in a plasma with $\gamma_\star\gg 1$ and a mildly relativistic $\gamma_{\rm peak}\sim 2$ in order to explain observations. Then, one can use the simplified estimate $\nu_{\rm em} \approx \gamma^{1/2}_\star\nu_p$. The values of $\nu_{\rm em}$ in our fiducial global magnetospheric model are shown in Figure~\ref{fig:power_freq}. 

The generated energy flux of radio emission in the RL frame is given by Eq.~(\ref{eq:power_RL}), which we rewrite as
\begin{equation}
S'_{\rm EM} = j'_0\,\frac{m_e c^2\alpha_{v'}}{e\,\omega'_p t'_{\rm cool}},
\qquad
\alpha_{v'}\equiv\frac{\beta'_d}{2}(\gamma'_d-1).
\end{equation}
The corresponding flux in the lab frame is $S_{\rm EM} =\gamma^2_\star S'_{\rm EM}$. Substituting $j_0' = \gamma_\star j_0$ (this relation holds because the plasma flow is nearly charge-neutral) and using Eq.~(\ref{eq:current_density}) for $j_0$, we obtain
\begin{equation}
    S_{\rm EM} = \frac{\gamma_\star^3 c\,eB\psi}{4\pi r_eR_{\max}}
    \frac{\alpha_{v'}}{\omega'_p t'_{\rm cool}}.
\end{equation}
The apparent luminosity of radio emission (isotropic equivalent) in the lab frame is $L_{\rm EM}\approx 4\pi r^2 S_{\rm EM}$, and we find
\begin{equation}
    L_{\rm EM}\sim 10^{30}\left(\frac{\gamma_\star}{10}\right)^3 \psi\alpha_{v'}\left(\frac{10^4}{\omega'_p t'_{\rm cool}}\right) \,{\rm erg~s}^{-1}.
    \label{eq:power_rest}
\end{equation}
Here, we assumed a magnetosphere with the dipole moment $\mu\approx 1.5\times 10^{32}$\,G\,cm$^3$, which is observed for the canonical radio magnetar XTE~J1810-197 and corresponds to the surface magnetic field at the star's pole $B_*\approx 3\times 10^{14}$\,G.

Bright radio emission is expected in the region where either electrons or positrons experience a strong radiative drag, $D_{\pm}>1$, since the two-stream instability is sustained only in regions of strong drag. The observed radio luminosity of XTE~J1810-197 also requires $\gamma_\star\sim 10$ and a sufficiently
low pair multiplicity ${\cal M}$, so that the $e^{\pm}$ beams are trans-relativistic in the RL frame, $\alpha_{v'}\sim 1$. These conditions are satisfied at radii $r\sim (15-50)R_*$ on closed magnetic field lines not very far from the magnetic axis, as shown in Figure~\ref{fig:power_freq} (which assumes ${\cal M}=50$).
Emission from this extended region spans a wide range of frequencies $\nu_{\rm em}$,  from one to a few tens of GHz. The significant extent of the emission region across magnetic field lines is also consistent with the broad radio pulses observed in magnetars.

As the radio waves propagate to the outer magnetosphere at $r>10^3R_*$, they can undergo cyclotron absorption. It
occurs when the wave frequency transformed to the particle rest frame matches the cyclotron frequency, $\gamma(\omega-{\boldsymbol{k}}\cdot {\boldsymbol{v}}) = \omega_B$. Cyclotron absorption is a significant problem for the escape of emission in ordinary radio pulsars \citep[e.g.,][]{Philippov_2022}. In Appendix~F, we show that in the case of magnetars, the resonance occurs at $r_{\rm abs}\sim {\rm several}\times 10^3 R_*$. We also find the absorption optical depth $\tau_{\rm abs}
\sim (1 - \cos\theta_c){\cal M}\sin^2\theta_c$ up to geometric factors of order unity (see Appendix~\ref{appen:cyclotron_absorb} for the full expression), where $\theta_c$ is the latitude of the resonance location. The optical depth is reduced near the axis of the twisted magnetosphere because its current density $j_0$ (Eq.~\ref{eq:current_density}) and the corresponding plasma density (Eq.~\ref{eq:number_density}) drop near the axis.
Radio waves produced at smaller radii will escape through the outer magnetosphere if they encounter $\tau_{\rm abs}<1$. This condition defines a polar ``visibility zone'' inside the emission region that produces escaping radiation. Its boundary is indicated by the blue curve in Figure~\ref{fig:power_freq}. As one can see, the region of bright radio emission (generated with $\gamma_\star\sim 10$) lies within the visibility zone, and therefore it is not blocked by cyclotron absorption at $r_{\rm abs}$.

Note that the values of $\gamma_\star\sim 10$ are found in the proposed region of radio emission when radiative drag is dominated by X-rays flowing radially from the magnetar, as assumed in Figure~\ref{fig:gamma}. A more advanced model involves additional drag created by X-rays scattered in the magnetosphere, which reduces $\gamma_\star$ \citep{Beloborodov_2013}. The contribution of scattered radiation depends on the anisotropy of magnetar's surface emission (it may be dominated by the heated footprint of the twisted ``$j$-bundle'' around the magnetic pole) and the distribution of magnetospheric electric currents, which are gradually erased in the inner magnetosphere \citep{Beloborodov_2009}. The complicated global picture will be further studied in future work.

\section{Summary and Discussion} \label{sec:summary}
Our first-principles simulations of the magnetospheric current in the radiatively locked state at radii $r \lesssim 50 R_*$ show the following:
\begin{itemize}
    \item 
    The radiative drag causes a sustained two-stream instability of the electron and positron flows that carry the electric current in the twisted magnetosphere at $r\sim 20-50R_*$. The instability saturates by driving strong turbulence that leads to particle trapping in the amplified electrostatic waves. In the resulting steady state, the broadening of the $e^\pm$ momentum distributions by turbulent electric fields balances the squeezing effect of the radiation drag.
    \item Our two-dimensional simulations (performed in the ``radiatively-locked'' reference frame) demonstrate that a fraction of the excited turbulence energy is channeled into electromagnetic fluctuations in the superluminal mode, which propagates freely through plasmas with decreasing density and forms the escaping radio emission. The emission power calculated for the conditions in the magnetar magnetosphere is sufficient to explain the observed luminosity of $\sim 10^{30}\, \mathrm{erg/s}$. Most of the power is concentrated around the relativistic plasma frequency of the counter-streaming hot electron-positron flow. The observed radiation is Doppler-shifted to higher frequencies due to the relativistic motion of the radiatively locked frame. This results in a broad range of emission frequencies, extending far above 1~GHz.
\end{itemize}

The radiative drag force serves as an engine that persistently drives the instability generating GHz radio waves. Therefore, the location of the radio-emitting region and the observed spectrum are determined by the properties of the radiation field in magnetar magnetospheres. Taking into account only thermal X-ray photons propagating radially from the magnetar, we find an extended emission region near the magnetic axis at radial distances ranging from 15 to 50 magnetar radii. This finding is consistent with the observed wide pulse profiles of radio emission. Future detailed calculations of the observed beam shape should take into account refraction, which is significant near the emission point where $\nu_{\rm em}\approx \nu_p$ \citep[e.g.,][]{LyubPet98}.

The highest frequency in the radio-emitting zone, as calculated from a global two-fluid model, falls somewhat short of the $100$\,GHz that is observed in radio magnetars \citep{Torne2015}. One possible explanation for this discrepancy is our simplistic assumption (adopted in the global model) that the observed emission
frequency is tightly coupled to the local plasma frequency. In more detailed models, the frequency will be increased: in the simulations performed in large periodic boxes, which allow substantially longer times for nonlinear interactions, we find that the spectrum of superluminal modes extends to at least $\approx 3\nu'_p$, pushing 
the predicted frequency cutoff towards $\approx 100\,{\rm GHz}$. We expect a similar increase of $\nu'_{\rm em}$ for our simulations with non-periodic boundaries (which allow the escape of electromagnetic waves)
when they are extended to larger simulation boxes. We will explore the mechanism of frequency increase in large boxes in future work.

Polarization of radio emission has been observed in the canonical
radio magnetar J1810--197 following outbursts in 2003 \citep{KramerStappers2007} and 2018 \citep{Dai2019}, as well as in magnetars J1550--5418 \citep{Camilo_2008} and J1818--1607 \citep{Fisher2024}. Strongly linearly polarized radiation was detected in all cases, consistent with the production of coherent radiation in the highly magnetized pair plasma in the magnetosphere. Key polarization characteristics, including the swing of the position angle of the linearly polarized component and the degree of circular polarization as a function of the pulse phase, exhibit diverse and time-evolving characteristics. Although not always resembling the S-shape behavior predicted by the classic ``rotating vector model'' (RVM), in most cases, the position angle shows a significant swing that is consistent with the origin in the polar flux tube described in our model. The wave normal modes in the inner magnetosphere are linearly polarized. Thus, the often observed significant circular polarization (e.g., $\approx15\%$ in J1550--5418 and 1818--1607 to nearly $50\%$ in parts of the profile of J1810--197 after the 2018 outburst) has to originate from propagation effects in the outer magnetosphere.  The observed linear
polarization of escaping radiation is also set at a large radius $r_{\rm pl}$ \citep[e.g.,][]{ChengRuderman1979,PetLyub2000,BeskinPh2012}. Inside this radius, the difference of refractive indices of the two modes $\Delta n$ is sufficiently large to keep the wave in the local eigen mode. Hence, its  polarization vector tracks the gradually changing direction of the local magnetic field $\boldsymbol{B}$, which may be described by angle $\phi_B$ in the plane perpendicular to the direction of propagation. The adiabatic tracking condition breaks at $r\sim r_{\rm pl}$ where $\Delta n \sim (\partial \phi_B/\partial r) c/\omega_{\rm em}$, and at larger radii the wave polarization vector ``freezes out,'' i.e. stops evolving along the ray. Evaluating this condition for typical parameters of plasma in magnetar magnetospheres gives $r_{\rm pl}/R_{*}\sim10^3-10^4$, comparable to the cyclotron resonance radius.\footnote{Explicit equations for $\Delta n$ and other characteristics of normal modes in the limiting polarization region are described in Appendix \ref{appen:polarization}.} The high altitude of the polarization-limiting radius is consistent with significant deviations from RVM behavior and the appearance of substantial circular polarization \citep{PetLyub2000,BeskinPh2012}, although more careful calculations utilizing a realistic spatial plasma density distribution are required to make quantitative statements.

To conclude, in this Letter, we established a model of magnetar radio emission based on first-principles plasma simulations. In the future, the availability of global, self-consistent kinetic models of plasma flow in the magnetospheres of magnetars can provide realistic distributions of the radiation field and plasma properties, such as the multiplicity and Lorentz factors of $e^\pm$ on different magnetic field lines. Such advanced models can be used to calculate the radio spectra and polarized light curves suitable for detailed comparisons with radio observations, an exciting possibility that we will explore in future work. 

We thank Roger Blandford, Ashley Bransgrove, Sasha Chernoglazov, Robbie Ewart, Hayk Hakobyan, Jens Mahlmann, Anatoly Spitkovsky, and Chris Thompson for fruitful discussions. This work was supported by NSF grant No. AST-2307395 (SZ and AP), Simons Foundation (00001470, AP), and facilitated by Multimessenger Plasma Physics Center (MPPC, AP and AB), NSF grant No. PHY-2206607. A.P. additionally acknowledges support by an Alfred P. Sloan Fellowship, and a Packard Foundation Fellowship in Science and Engineering. Computing resources were provided and supported by the Division of Information Technology at the University of Maryland (\href{http://hpcc.umd.edu}{\texttt{Zaratan} cluster}\footnote{\url{http://hpcc.umd.edu}}). This research is part of the Frontera computing project \citep{Frontera} at the Texas Advanced Computing Center (LRAC-AST21006). Frontera is made possible by NSF award OAC-1818253. This research was additionally supported in part by grant NSF PHY-2309135 to the Kavli Institute for Theoretical Physics (KITP).

\appendix

\section{Two-stream Instability threshold for Lorentzian distributions} \label{appen:stability_threshold}
Criteria for the stability of hot counter-streaming $e^{\pm}$ beams can be computed exactly for the Lorentzian non-relativistic distribution function \citep[e.g.,][]{ONeil_1968}, which reads
\begin{equation}
    f_\pm (v) = \frac{n v_T}{\pi} \frac{1}{(v \pm v_d)^2 + v_T^2},
\end{equation}
where $v_T$ is the thermal spread of the distribution. The dielectric function, $\epsilon$, can be computed exactly using the residue theorem as
\begin{align}
    \epsilon & = 1 + \frac{4\pi e^2}{k^2} \sum_{s=\pm} \int \mathrm{d}^3\boldsymbol{p} \frac{\boldsymbol{k} \cdot \nabla_p f_s}{\omega - \boldsymbol{k} \cdot \boldsymbol{v}} \notag \\
    &= 1 - \omega_p^2 \left(\frac{1}{(\omega - k v_d + \mathrm{i}k v_T)^2} + \frac{1}{(\omega + k v_d + \mathrm{i}k v_T)^2} \right).
\end{align}
The dispersion relation of 1D electrostatic modes is given by $\epsilon = 0$, 
\begin{equation}
    (\omega + \mathrm{i} k v_T)^2 = (k^2 v_d^2 + \omega_p^2) \pm \sqrt{\omega_p^2 (\omega_p^2 + 4k^2 v_d^2)}.
\end{equation}
The growth rate of the potentially unstable mode is
\begin{equation}
    \Gamma = \sqrt{\sqrt{\omega_p^2 (\omega_p^2 + 4k^2 v_d^2)} - (k^2 v_d^2 + \omega_p^2)} - k v_T, 
\end{equation}
and $\Gamma >0$ yields
\begin{equation}
    k^2 (v_T^2 + v_d^2) + 2\omega_p^2 (v_T^2 - v_d^2) < 0.
\end{equation}
If $v_T \geq v_d$, the above inequality is not satisfied for any value of $k$, indicating that the system is stable. If $v_T < v_d$, the system is unstable for $k^2 < 2\omega_p^2 (v_d^2 - v_T^2) / (v_T^2 + v_d^2)^2$. Thus, the two-stream instability is triggered when the drift velocity, $v_d$, exceeds the thermal spread of a single beam, $v_T$.

\section{Prediction of saturated electric field energy: a quasi-linear argument} \label{appen:scaling_quasi_linear}
As mentioned in the main text, instability saturates through particle trapping originating from large kicks due to highly localized intense electric field fluctuations. It turns out that a similar scaling for the saturated electric field energy density can be derived using quasi-linear theory, which assumes a simplified diffusive picture with random, small-amplitude kicks. In the quasi-linear approach, the saturated state involves a balance between diffusion in momentum space due to turbulent electric fields and the squeezing of the distribution functions of $e^{\pm}$ by the drag force. Suppose that particles experience a stochastic electric field that changes sign on a timescale $\Delta t \sim 1/\omega_p$, leading to a diffusion in the momentum space with diffusion coefficient 
\begin{equation}
    D_{pp} \sim \frac{(eE_x \Delta t)^2}{\Delta t} \sim
    \frac{(eE_x)^2}{\omega_p}.
\end{equation}
In the saturation state, the diffusion should be balanced by the effect caused by the drag force, 
\begin{equation}
    D_{pp} \frac{\partial f}{\partial p} \sim (q E_0 + F_{\rm rad}) f.
\end{equation}
Inserting $(q E_0+F_{\rm rad}) \sim m_e v_d/t_{\rm cool}$, and $f/(\partial f/\partial p) \sim m_e v_d$, one obtains an estimate of the steady-state energy density of the stochastic field, 
\begin{equation}
    \epsilon_E = \frac{E_x^2/8\pi}{{\cal E}_K} \sim \frac{1}{\omega_p t_{\rm cool}},
    \label{eq:saturated_E_sq}
\end{equation}
where ${\cal E}_K = nm_ev_d^2/2$ is the kinetic energy density of the plasmas.

\section{Synchrotron Cooling Strength and Possibility of 1D Transverse Electromagnetic Instabilities in the Outer Magnetosphere}\label{app:transverse_instab}

In the region where radiation drag due to resonant inverse Compton scattering slows down the plasma and the two-stream instability gets excited, synchrotron losses happen almost instantaneously. Thus, particle motion can be safely assumed to be one-dimensional, i.e., aligned with the background magnetic field. In principle, the production of coherent emission could also occur in the outer magnetosphere, where synchrotron cooling is weak, and the excitation of currents across magnetic field lines becomes possible. In this Appendix, we estimate the timescale of synchrotron losses and consider whether any electromagnetic instabilities in counter-streaming plasma are possible when transverse Larmor motion is allowed. We first establish that synchrotron cooling becomes slow at $r\gtrsim 10^3R_*$, and then calculate the wave dispersion relation and instabilities in the outer zone.

The strength of the synchrotron cooling can be parametrized by comparing the timescale of synchrotron losses $t_{\rm syn}$ and local light crossing time, 
$t_{\rm syn}/(r/c) = 3 / (4\gamma \mathscr{L}_B \sin^2\chi)$, where
\begin{equation}
   \mathscr{L}_B \equiv \frac{U_B\sigma_T r}{m_e c^2} = 0.32 \times \bigg(\frac{B}{10^5\mathrm{G}}\bigg)^2 \bigg(\frac{r}{1000 R_*}\bigg)
    \label{eq:syn_time}
\end{equation}
is the magnetic compactness parameter, $\chi$ is the particle's pitch angle, $U_B=B^2/8\pi$ is the energy density of magnetic field, $\sigma_T$ is the Thomson cross section. In the inner magnetosphere, $r\lesssim 50 R_*$, synchrotron losses occur on a timescale much shorter than the timescale of the radiative drag, $t_{\rm syn} \lesssim 10^{-7} (r/c) \sim 10^{-6} t_{\rm cool}$, and can therefore be considered instantaneous. The synchrotron cooling timescale becomes comparable to the light crossing time at $r\sim 10^3R_*$ due to the rapid, $\propto r^{-3}$, decay of the magnetic field strength. 

Beyond $\sim 100R_*$, radiative drag becomes weak, while the two-stream instability at smaller radii substantially broadened the distribution functions of electrons and positrons. Thus, we expect no further growth of electrostatic instabilities. However, a non-zero velocity difference between electrons and positrons is still required to sustain the current, and other streaming instabilities may arise when synchrotron losses weaken. 

To explore the possible instabilities resulting from allowing the transverse motion in 1D, we consider wave dispersion relation for two cold, counter-streaming beams allowing finite gyroradii, with a wavevector $\boldsymbol{k}$ parallel to the background field $\boldsymbol{B}_0$. The susceptibility tensor, $\chi_{ij}$, of a single cold beam streaming with drift velocity $v_d \hat{x}$ is 
\begin{equation}
    \begin{pmatrix}
    \dfrac{k_\perp^2 v_d^2 \omega_p^2}{\gamma_d \omega^2 \tilde{\omega}_B^2} - \dfrac{\omega_p^2}{\gamma_d^3 \tilde{\omega}^2} & \dfrac{k_\perp v_d \tilde{\omega} \omega_p^2}{\gamma_d \omega^2 \tilde{\omega}_B^2} & \dfrac{\mathrm{i}k_\perp v_d \omega_B \omega_p^2}{\gamma_d^2 \omega^2 \tilde{\omega}_B^2} \\[10pt]
    \dfrac{k_\perp v_d \tilde{\omega} \omega_p^2}{\gamma_d \omega^2 \tilde{\omega}_B^2} & \dfrac{\omega_p^2 \tilde{\omega}^2}{\gamma_d \omega^2 \tilde{\omega}_B^2} & \dfrac{\mathrm{i} \omega_B \tilde{\omega} \omega_p^2}{\gamma_d^2 \omega^2 \tilde{\omega}_B^2} \\[10pt]
    -\dfrac{\mathrm{i}k_\perp v_d \omega_B \omega_p^2}{\gamma_d^2 \omega^2 \tilde{\omega}_B^2} & \dfrac{-\mathrm{i} \omega_B \tilde{\omega} \omega_p^2}{\gamma_d^2 \omega^2 \tilde{\omega}_B^2} & \dfrac{\omega_p^2 \tilde{\omega}^2}{\gamma_d \omega^2 \tilde{\omega}_B^2}
    \end{pmatrix},\\
    \label{eq:susceptibility}
\end{equation}
where $\tilde{\omega} \equiv (\omega - k_\parallel v_d), \; \tilde{\omega}_B^2 = (\omega_B^2 / \gamma_d^2 - \tilde{\omega}^2)$, and the wave vector $\boldsymbol{k}$ lies in the $x-y$ plane, $\boldsymbol{k} = k_\parallel \hat{x} + k_\perp \hat{y}$, as in our simulations. For parallel propagating modes, $k_\perp = 0$, the nonzero components of the dielectric tensor are
\begin{align}
    \epsilon_{xx} &= 1 - \frac{\omega_p^2}{\gamma_d^3}\sum_{\pm}\frac{1}{(\omega \pm k v_d)^2}, \notag \\
    \epsilon_{yy} &= \epsilon_{zz} = 1 - \frac{\omega_p^2}{\gamma_d \omega^2} \sum_{\pm}\frac{(\omega \pm k v_d)^2}{(\omega_B / \gamma_d)^2 - (\omega \pm k v_d)^2}, \notag \\
    \epsilon_{yz} &= -\epsilon_{zy} = \frac{\mathrm{i}|\omega_B|\omega_p^2}{\gamma_d^2 \omega^2} \sum_{\pm} \pm \frac{\omega \pm k v_d}{(\omega_B / \gamma_d)^2 - (\omega \pm k v_d)^2}.
\end{align}
The wave equation is $(c^2/\omega^2) \boldsymbol{k} \times (\boldsymbol{k} \times \boldsymbol{E}) + \boldsymbol{\epsilon} \cdot \boldsymbol{E} = 0$, and the dispersion relation is given by
\begin{align}
    \det &\bigg[\boldsymbol{\epsilon} - \frac{k^2 c^2}{\omega^2} \boldsymbol{I} + \frac{c^2}{\omega^2} \boldsymbol{k} \boldsymbol{k}\bigg] \notag \\
    &= \epsilon_{xx} \big[(\epsilon_{yy} - \frac{k^2 c^2}{\omega^2})(\epsilon_{zz} - \frac{k^2 c^2}{\omega^2}) - \epsilon_{yz} \epsilon_{zy}\big] = 0.
\end{align}
It is straightforward to see that $\epsilon_{xx} = 0$ gives the electrostatic two-stream instability, while the transverse electromagnetic modes are given by $\big[(\epsilon_{yy} - k^2 c^2/\omega^2)(\epsilon_{zz} - k^2 c^2/\omega^2) - \epsilon_{yz} \epsilon_{zy}\big] = 0$. The dispersion relation can be written as the product of two cubic equations of $\omega^2$, and one of its six modes is unstable. In the magnetosphere, the magnetization is high even at $r\sim 1000 R_*$,
\begin{equation}
    \sigma = 6\times 10^8 \left(\frac{r}{1000R_*}\right)^{-3} \left(\frac{{\cal M}}{100}\right)^{-1} \left(\frac{R_{\rm max}}{1000R_*}\right) \left(\frac{B_*}{10^{14}\mathrm{G}}\right).
\end{equation}
In the $\sigma \gg 1$ limit, the mode is unstable in two separate k-ranges: $0 < k < 2 v_d \omega_p^2 / (|\omega_B| c^2)$, and $(1 - 2\beta_d^2 \gamma_d / \sigma) ((|\omega_B| / \gamma_d)/v_d) < k < (|\omega_B| / \gamma_d) / v_d$. The corresponding maximum growth rates are $\beta_d \omega_p / \sqrt{\sigma}$ and $\beta_d^2 \omega_p / \sqrt{\sigma}$, respectively.

\cite{lyutikov2020} considered the same dispersion relation and identified unstable branches. Their interpretation of the instability near $k = 0$ as a firehose instability, however, is not correct. In fact, by rewriting the instability threshold as $k < 8\pi n e v_d/(Bc) = 4\pi j_0 / (Bc)$, it is clear that this is a kink-like instability relevant only on large spatial scales, $\lambda > Bc/(2j_0) \approx (2\pi/\psi) R_{\mathrm{max}}$, and is therefore irrelevant for quiescent stable magnetospheres. 

The instability near $k \sim (|\omega_B| / \gamma_d) / v_d$ is the gyro-resonance instability, which is active in a narrow range of wavenumbers, $\Delta k/k \sim 2\beta^2_d\gamma_d/\sigma \ll 1$. Furthermore, the timescale of its growth, $\sqrt{\sigma}/ (\beta_d^2 \omega_p) \sim 2 R_{\rm max} / (\beta_d c)$, is macroscopic. We expect the growth rate of the instability to be even smaller for more realistic hot beams. 

To conclude, we do not expect any of the transverse instabilities to play a role in powering the observed radio emission. In comparison, the timescale of the radiatively driven two-stream instability is comparable to the cooling time, which is smaller than the macroscopic timescale by a factor of $D$. 

\section{Oblique Two-stream unstable modes in the inner magnetosphere} \label{appen:eigenmodes}
In this Appendix, we calculate the properties of two-stream unstable modes allowing finite angle of propagation with respect to the magnetic field. At $r \lesssim 50 R_*$, where radiatively driven two-stream instability operates, it is appropriate to treat the field strength as infinite. In this limit, the only non-vanishing component of the susceptibility tensor (see Eq.~\ref{eq:susceptibility}) is $\chi_{xx} = -\sum_{\pm} \omega_p^2/(\gamma_d^3 (\omega \pm k_\parallel v_d)^2)$.

The dispersion relation of the ordinary mode (with non-zero $E_x, E_y, B_z$) is 
\begin{align}
    \hat{\omega}^6 &- \left(2\hat{k}_\parallel^2 \beta_d^2 + \hat{k}^2 + \frac{2}{\gamma_d^3}\right)\hat{\omega}^4 \notag \\
    &+ \hat{k}_\parallel^2\left(\hat{k}_\parallel^2 \beta_d^4 + 2\hat{k}^2 \beta_d^2 + \frac{2}{\gamma_d^5}\right) \hat{\omega}^2 + \hat{k}_\parallel^4 \beta_d^2 \left(\frac{2}{\gamma_d^3} - \hat{k}^2 \beta_d^2\right) = 0 \notag,
\end{align}
where $\hat{\omega}\equiv \omega/\omega_p, \hat{k} \equiv kd_e$, and $\hat{k}_{\parallel}$ is the field-aligned component of the wavenumber. This is a cubic equation for $\hat{\omega}^2$, and its three solutions correspond to the superluminal mode, the Alfven mode, and the linearly unstable mode. The threshold of the instability is determined by $\hat{\omega}^2 = 0$, which gives $\hat{k}^2 = 2 / (\gamma_d^3 \beta_d^2)$, independent of the propagation angle. For the unstable mode, one has $|\hat{\omega}^2| \leq \hat{\Gamma}_{\rm max, 0}^2 = 1 / (4\gamma_d^3) \ll 1$, where $\hat{\Gamma}_{\rm max, 0} = 1/(2\gamma_d^{3/2})$ is the normalized maximum growth rate for electrostatic modes propagating exactly along the background magnetic field. Therefore, an approximate form of the dispersion relation of the unstable mode may be obtained by neglecting the terms of order higher than $\hat{\omega}^2$,
\begin{equation}
    \hat{\Gamma}^2 \approx \frac{\hat{k}_\parallel^2 \beta_d^2(2/\gamma_d^3 - \hat{k}^2\beta_d^2)}{\hat{k}_\parallel^2 \beta_d^4 + 2\hat{k}^2\beta_d^2 + 2/\gamma_d^5}.
    \label{eq:growth_rate}
\end{equation}
Polarization of the unstable eigenmode is determined by $c \nabla_x \times B_z = -\partial_t E_y$ and $c(\partial_xE_y - \partial_y E_x) = -\partial_tB_z$, which leads to 
\begin{equation}
    |B_z| = \left |\frac{\hat{\Gamma}}{\hat{k}_\parallel} E_y \right | = \frac{\hat{\Gamma} \hat{k}_\perp}{\sqrt{\hat{k}_\perp^2 \hat{k}_\parallel^2 + (\hat{\Gamma}^2 + \hat{k}_\parallel^2)^2}} |E|,
    \label{eq:B_to_E}
\end{equation}
where $|E| = \sqrt{|E_x|^2 + |E_y|^2}$. When $\beta_d \lesssim 0.5$, we may neglect the term $\hat{k}_\parallel^2 \beta_d^4$ in the denominator in Eq.~(\ref{eq:growth_rate}), leading to $\hat{\Gamma} \propto \cos\theta_k$. The normalized maximum growth rate is at least several times smaller than the corresponding normalized wave vector, $\hat{\Gamma}_{\rm max} \sim (\cos\theta_k)/2 < \sqrt{3}/(2\beta_d) \sim \hat{k}_{\rm max}$. Thus, one may neglect the term $\hat{\Gamma}^2$ in the denominator in Eq.~(\ref{eq:B_to_E}), leading to $|B_z/E| \sim \hat{\Gamma} \tan\theta_k / \hat{k}$. For the fastest-growing mode, $|B_z/E| \sim \beta_d\sin\theta_k/\sqrt{3}$.

\section{Propagation of superluminal waves in 2X1V simulations} \label{appen:superluminal_propagation}
\begin{figure*}[ht!]
\includegraphics[width=18cm]{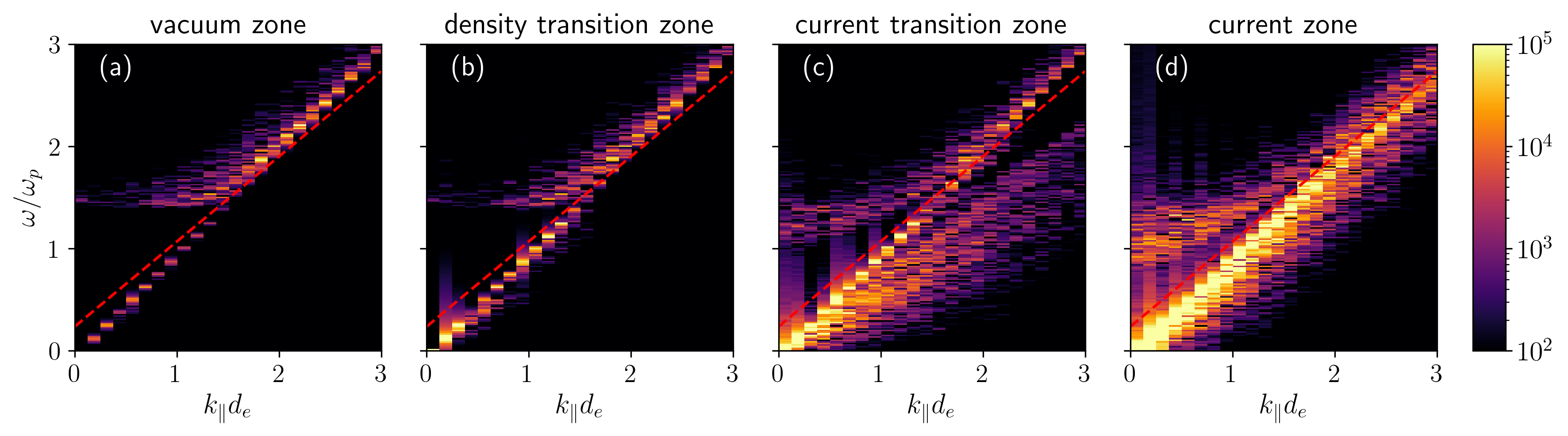}
\caption{Fourier intensities of the magnetic field, $|B_z(\omega,{\boldsymbol{k}})|^2$, calculated in four slices in different zones in a simulation with $v_d/c=0.5$, $\omega_pt_{\rm cool}=3000$. The regions above the red dashed lines are defined as the superluminal part of the spectrum.
\label{fig:superluminal_propagation}}
\end{figure*}
\begin{figure}[ht!]
\includegraphics[width=\linewidth]{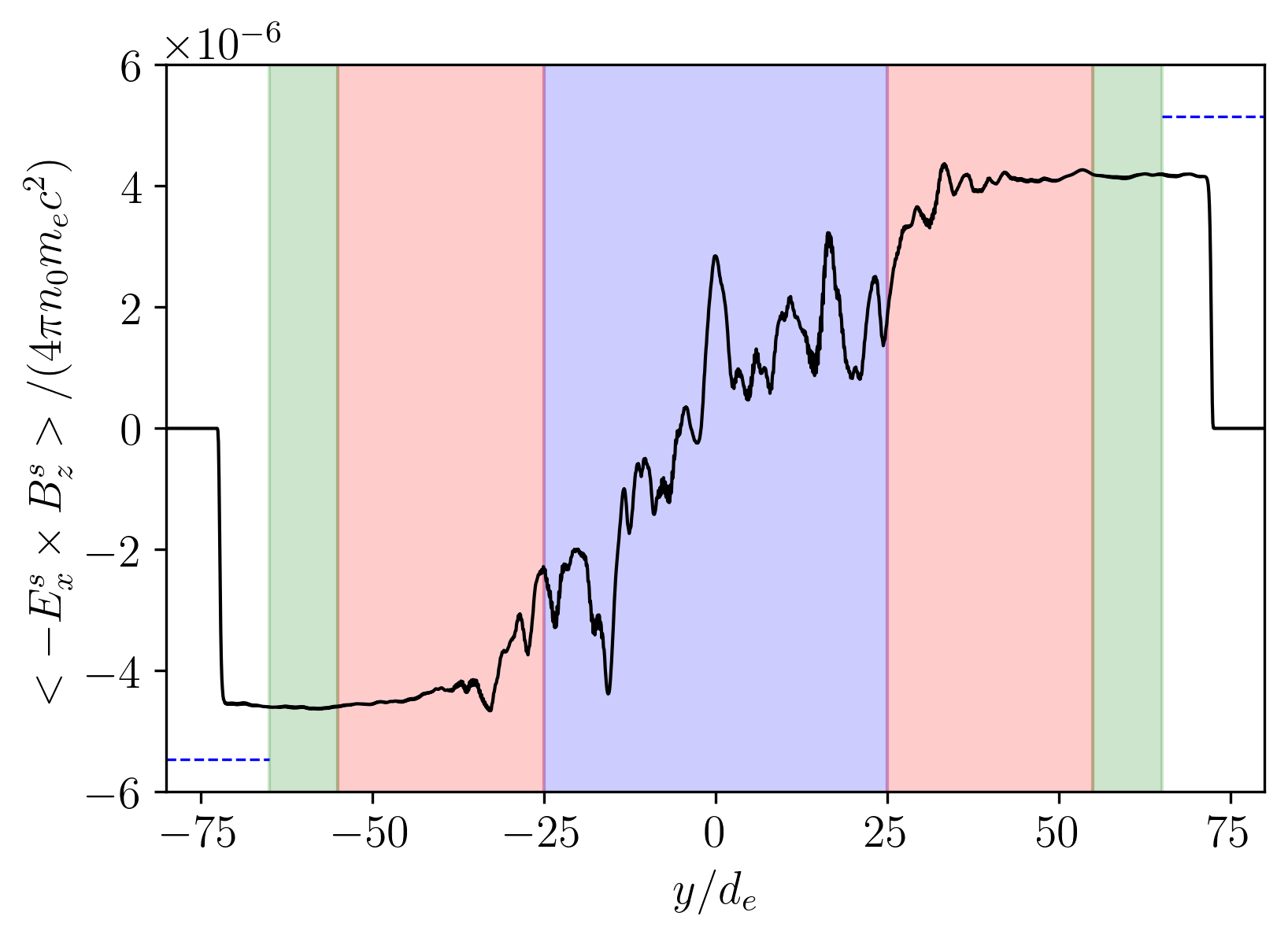}
\caption{Poynting flux of the superluminal component of the electromagnetic field in a simulation with $v_d/c=0.5$, $\omega_pt_{\rm cool}=3000$. The Poynting flux is computed from the reconstructed spatial field components, which are obtained by performing an inverse Fourier transform for the superluminal part of the spectrum. The purple, red, green, and white colored regions show the central current zone, current transition zone, density transition zone, and vacuum zone, respectively. The values of Poynting fluxes computed from the full electromagnetic field components in the vacuum zone are shown by the blue dashed lines. 
\label{fig:superluminal_poynting}}
\end{figure}
To verify that the electromagnetic waves measured in the vacuum zone indeed correspond to superluminal waves that are generated in the central current zone, we perform spatial, i.e., along x, and temporal Fourier transform on every horizontal (constant-$y$) slice. The resulting intensities in Fourier ($\omega-k_\parallel$) space for slices in each zone are shown in Figure~\ref{fig:superluminal_propagation}, which clearly demonstrates that the superluminal mode propagates through plasmas without being damped. To reconstruct spatial components of the fields corresponding to superluminal waves, we further perform an inverse Fourier transform for the superluminal part of the spectrum, defined as the region beyond the red dashed lines in Figure~\ref{fig:superluminal_propagation}, which well separate the superluminal mode and the subluminal mode. Figure~\ref{fig:superluminal_poynting} shows the Poynting flux computed from these reconstructed fields, which clearly indicates that the superluminal waves are generated in the central current (purple) zone. They propagate through the current transition (red) zone and the density transition (green) zone without any significant damping and enter the vacuum (white) zone, forming the electromagnetic waves we measure near the absorbing layers. The Poynting flux computed from the full electromagnetic fields in the vacuum zone (blue dashed lines) is very close to the Poynting flux calculated from the reconstructed fields.

\section{Optical Depth of cyclotron absorption} \label{appen:cyclotron_absorb}
The cyclotron absorption can occur when the frequency of the electromagnetic waves seen by the particles matches with the cyclotron frequency, $\gamma \tilde{\omega} = \gamma (1 - \beta \cos \phi) \omega = \omega_B$. Taking $\nu \sim 10 \mathrm{GHz}$ and assuming that the effect of the Doppler shift is of order unity, it can be estimated that the cyclotron resonance occurs at $r_c \sim $ several $\times 10^3 R_*$. Since the resonance occurs far from the magnetar, we can make the approximation that the emission propagates radially. The cross-section for cyclotron absorption is $\sigma_{\rm cyc} = 2\pi^2 e^2 / (m_e c) \delta(\gamma \tilde{\omega} - \omega_B)$, and the optical depth of the radiation emitted in the direction along the field line, at $(r_0, \theta_0)$, is 
\begin{align}
    &\tau_{\rm cyc} (r_0, \theta_0) = \int_{r_0}^{+\infty} n \sigma_{\rm cyc} (1 - \beta \cos \phi) \mathrm{d}r \notag \\
    &\;\approx \frac{2\pi^2 e^2}{m_e c} \frac{n (1 - \beta \cos \phi)}{|\mathrm{d}\omega_B / \mathrm{d}r|} \bigg|_{r = r_c} = \frac{\pi}{6}\frac{(1 - \beta_c \cos\phi_c)}{\beta_c} {\cal M} \sin^2 \theta_c, 
\end{align}
where $\beta_c, \phi_c, \theta_c$ are the particle velocity, the angle between the wave propagation and the particle velocity, and the coordinate at the location where the resonance occurs. Since emission is beamed along the field line, one can estimate $\theta_c \approx \theta_0 + \tan^{-1}(\tan\theta_0 / 2)$, and $\cos\phi_c = 2 \cos\theta_c / \sqrt{1 + 3\cos^2\theta_c}$. Since the plasma flow remains mildly relativistic as it streams out, we assume $\beta_c \approx 1$.

\section{Polarization of Normal modes} \label{appen:polarization}
Polarization of normal modes at large altitudes where radio emission decouples from the plasma can be calculated by solving the dispersion relation for the dielectric tensor corresponding to Eq.~(\ref{eq:susceptibility}). The difference between the refractive indexes of two electromagnetic normal modes can be expressed as \citep[e.g.,][]{ChengRuderman1979,BeskinPh2012}:

\begin{align}
\Delta n \approx\frac{1}{2}\frac{\omega^2_p\omega^2_B}{\gamma^3\tilde{\omega}^2(\omega^2_B-\gamma^2\tilde{\omega}^2)} \frac{\sqrt{q^2+1}}{q}\sin^2\phi,
\label{eq:deltan}
\end{align}
where parameter $q\approx i(\epsilon_{\rm y'y'}-\epsilon_{\rm x'x'})/2\epsilon_{\rm x'y'}$ describes wave polarization: $q\gg 1$ corresponds to linear polarization, while $q\ll 1$ describes circularly polarized waves. Here, components of the dielectric tensor, $\epsilon_{\rm i'j'}$, are computed in a coordinate system in which the $z'$ axis is directed along the wave vector, $\textbf{k}$, and the background magnetic field lies in the $x'z'$ plane. Direct calculation gives:
\begin{equation}
q\approx \frac{\omega_B {\cal M}}{2\omega\gamma^3(1-\cos\phi v_d/c)^2}\sin^2\phi.
\end{equation}

\bibliography{sample631}{}
\bibliographystyle{aasjournal}

\end{document}